**Origin of Pressure-Induced Structural Instability in CsPbX$_3$ Photovoltaic Perovskites**

*Sizhan Liu, Sandun Amarasinghe, Mo Li, Stella Chariton, Vitali Prakapenka, Sanjit K. Ghose, Yong Yan, Joshua Young[*] and Trevor A. Tyson[*]*

Sizhan Liu, Sandun Amarasinghe, Trevor A. Tyson

Department of Physics, New Jersey Institute of Technology, Newark, NJ 07102, USA

E-mail: tyson@njit.edu

Mo Li, Joshua Yong

Department of Chemical and Materials Engineering, New Jersey Institute of Technology, Newark, NJ 07102, USA

E-mail: jyoung@njit.edu

Stella Chariton, Vitali Prakapenka

Center for Advanced Radiation Sources, The University of Chicago, Argonne, IL 60439, USA

Sanjit K. Ghose

National Synchrotron Light Source II, Brookhaven National Laboratory, Upton, NY 11973, USA

Yong Yan

Department of Chemistry and Biochemistry, San Diego State University, San Diego, CA, 92182, USA

Under external stimuli, lead halide perovskites exhibit large atomic fluctuations, impacting optical and electron transport properties that affect device performance in operational settings. However, a thorough understanding of the atomic basis for the underlying structural instability is still absent. Focusing on the model material CsPbBr$_3$, the inherent lattice softness of halide perovskites is elucidated at the atomic level through *in-situ* single-crystal X-ray diffraction measurements under pressure complemented by atomic level simulations. We identify and explore the nature of the first-order phase transition to a distorted *P*2$_1$/*c* phase at 1.3 GPa, induced by the sudden Cs-Br bonds breaking. Unlike classical transition metal oxide perovskites, where the internal energy term dominates, we show explicitly that pressure primarily influences the Gibbs free energy for halide perovskites through the pressure-volume term. As such, strategically mitigating bond strains from volume shrinkage is the key to suppressing the first-order behavior for maintaining the coordinates of PbX$_6$ polyhedral upon external perturbation. Our thermodynamic calculation reveals the demand for high entropy in the $-T\Delta S$ term, which can be achieved by exploring a broader spectrum of doped A site and B sites in ABX$_3$ systems, enabling continuous structural changes that facilitate recovery from mechanical damage in practical applications.

# 1. Introduction

The structural stability of lead halide perovskites (LHPs) in response to external stimuli dictates their performance in photovoltaics,[1-2] optoelectronics,[3-4] and thermoelectric applications.[5-6] These materials have the chemical formula $APbX_3$, where A is an organic or inorganic cation, such as methylammonium ($MA^+$), formamidinium ($FA^+$), or $Cs^+$, and *X* represents a halide (I, Br, or Cl). The turnability of the three-dimensional perovskite structure principally depends on the cation in the A site, situated in the cuboctahedral cage of $PbX_6^{4-}$ octahedra units. This configuration endows inorganic LHPs (iLHPs) with improved thermal and chemical stabilities,[7-9] with power conversion efficiencies,[10] charge-carrier mobilities[11-12] and tunable band gap[13] compared to their hybrid analogues. Unfortunately, obtaining a clear picture of the behavior of these materials has proved challenging, partly because all the above-mentioned properties are affected by anharmonic atomic fluctuations,[14-16] leading to intrinsic structural instabilities, such as unforeseen structure transitions triggered by subtle temperature or stress fluctuations[17-18] and photo-degradation under irradiation and decomposition into photo-inactive phases[19-21] that hinder its commercial adoption. Substantial efforts have been devoted to understanding correlations between the atomic structure and physical properties of LHPs operating under complex conditions (*e.g.*, high temperatures, continuous light illumination, electric bias, or their combination)[22-27] in order to improve their performance. However, in-depth knowledge of these materials' intrinsic crystallographic structure and the formation of long-range defective lattice structures is absent. In particular, there are outstanding questions regarding how and what structures will be induced under mechanical pressure and thermal strains, which is postulated as the most typical external influence among various degradation mechanisms restricting the operational lifetime of the flexible perovskite solar cells (f-PSCs).[28-35]

Elucidating the interplay between these materials' ambient crystal structure and morphology upon deformation by external stimuli is critical to stabilizing LHPs under induced stress and strain. Recent investigations have unveiled discrepancies in the structures of these materials, revealing that the former symmetry assignments ascertained from powder diffraction data were not sufficiently precise.[18] High-pressure X-ray diffraction (XRD) analyses on powder $CsPbBr_3$ specimens have identified successive phase transitions at 1.3±0.2 and 2.4±0.2 GPa, respectively.[18, 36-37] The fact that these transitions occur at unexpectedly low pressures prompts a profound question on the intrinsic instability of LHPs, even though $CsPbBr_3$ is believed to be the most stable member of the APbX$_3$ system as photovoltaics.[9, 38] The Goldschmidt tolerance (*t*) of CsPbBr$_3$ (*t* =0.86)[39-42] and the anharmonic interactions, owing to the double-well nature of the energy potential in iLHPs,[15] indicate thermodynamic stability and facile lattice distortions such as tilting, rotation and volumetric change of the octahedral cages. The common denominator of these distortion modes is that they all arise due to the phase transition as a collective structural response to external perturbations. Such distortions break the symmetry in long-range order, leading to the emergence of half-integer reflections and resulting in a larger unit cell volume, as can be directly probed in the reciprocal space through single-crystal X-ray diffraction.

Previous high-pressure studies focused on polycrystalline samples, which averaged over multiple grains and failed to resolve weak diffraction signals corresponding to subtle symmetry breakings. These symmetry-breaking peaks overlap the strong high symmetry peaks. Therefore, we conducted synchrotron single-crystal X-ray diffraction measurements to elucidate the structural dynamics of halide perovskite, as it allows for *in-situ* characterization of single-grain crystallites with high-resolution in reciprocal space, surpassing previous powder-based techniques. Applying hydrostatic pressures using diamond anvil cells allows the mechanical stresses encountered in actual conditions to be replicated, revealing the correlations between stress-induced structure response and physical properties with the complement of density functional theory (DFT) calculations. Our single-crystal X-ray data revealed the low-pressure-induced pathway enabling facile structural transitions via progressive distortion of the PbBr$_6$ and CsBr$_{12}$ polyhedra, resulting in a tripled unit cell volume to accommodate the local symmetry reductions. These unique atomic dynamics unveil the deformation mechanism of the intrinsic lattice in response to stress. We rationalized our observations with DFT simulations, comparing the variations in internal energy against the *P*Δ*V* term of the Gibbs free energy, revealing the dominance of the *P*Δ*V* term concomitant with polyhedral distortion — a phenomenon that sharply contrasts with the behavior seen in complex oxide perovskites. Our DFT calculations also capture a large ~ 500 meV increase in the band gap across the 1.3 GPa phase transition, revealing a direct impact on the electronic and photovoltaic properties. The results enable a quantitative approach to access the interplay of stress-related structural degradation and the softness of Pb-X bonds in undoped APbX$_3$ systems, underpinning the necessity of circumventing the low-pressure structural transitions for designing high-performance perovskite-based devices.

## 2. Results

**2.1 High-Pressure Single Crystal X-Ray Diffraction Measurements**

High-pressure single-crystal X-ray diffraction analyses were performed on single-domain CsPbBr$_3$ crystal using diamond anvil cells to explore the compression-induced structural transitions from ambient condition to pressure up to 10.6 GPa, as illustrated in Figure 1. The crystallographic structures under pressure were revisited with more accurate atomic positions and appropriate space group assignments, including the subtle symmetry breakings overlooked in previous powder-based studies due to their weak scattering intensities. Notably, the structural transitions were observed at a low pressure of 1.3 GPa and a second transition between 4.7 and 5.6 GPa.

To reveal the emerging structural distortions that give rise to weak-intensity patterns in the reciprocal lattice upon compression, we first analyze the reciprocal space images to establish the accurate ambient CsPbBr$_3$ structure. In Figure 2a-c, the observed reciprocal lattice aligns with the simulated image of the solved monoclinic *P*2$_1$/*m* space group derived from all reflections found at 0.2 GPa, reproducing our previous work[18] and consistent with early structure studies;[43-44] however, ignoring the weak Bragg peaks yields the widely cited orthorhombic *Pnma* space group originally assigned in 1970s.[45] As pressure increases, polyhedral distortion and cation off-centering induce a reduced structural symmetry, resulting in a transition to a *P*2$_1$/*c* monoclinic structure when pressure exceeds 1.3 GPa ( Figure 2f). A comparison of the simulated and observed reciprocal lattice image at 1.6 GPa is displayed in Figure 2d-e, revealing that the distortions along [110] crystallographic direction of the ambient structure results in an elongated *c*-axis within the high-pressure cell (Figure 2f). As the pressure goes beyond 4.7 GPa, accumulated lattice distortions induce a subsequent symmetry reduction, evidenced by the emerging superlattice reflections and resulting in further unit cell volume expansion (Figure S7 to S9). These observed transitions of the crystalline structures through synchrotron single-crystal X-ray diffraction measurements challenge the previously postulated amorphous transition at 2.4 GPa[36-37] and call for a reevaluation of the currently accepted ambient structure.

To quantify the volume contraction induced internal energy change during phase transition, the Murnaghan equation of state (EOS), $V(P) = V_0 \left[ P\left(\frac{K'}{K}\right) + 1 \right]^{-1/K'}$, is applied to fit the pressure-dependent unit cell volume per CsPbBr$_3$ molecular unit (V/Z or volume per formula unit, f.u.) between 1.3 GPa and 4.7 GPa, similar to the approach employed for the classical RMnO$_3$ transition metal perovskite manganite.[46-47] The obtained bulk modulus $K$ and its pressure derivative $K'$ yield $6 \pm 2$ GPa and $9 \pm 2$ GPa, respectively. In Figure 3a, extrapolation of this model to ambient pressure reveals a $P\Delta V$ value of -0.076 eV/f.u. at the transition of 1.3 GPa, associated with a 5% volume change. (These results can be compared with measurements on the RMnO$_3$ (R=DY, Ho, Lu) perovskite system, which also exhibits large K' values indicative of anisotropic compression.[47]) To complement these experimental findings, DFT modeling is performed to offer insights into the energy landscape for the emerging superlattice upon pressurization, especially for the low-pressure-induced symmetry reduction to the $P2_1/c$ phase.

**2.2 Density Functional Theory Modeling**

The pressure-induced symmetry change arises from the collective response of atomic displacements, navigated by a free-energy landscape toward the more stable form of the competing phases. Mapping this free-energy landscape establishes a conceptual understanding of the inherent structure instability upon external compression. Here, we perform DFT modeling to elucidate the structural stability by examining the variations in internal energy as a function of pressure. Structural optimization is carried out for lattice parameters and atomic positions utilizing local density approximation (LDA) exchange functionals guided by experimentally derived crystallographic results (Table S2 to S5). We first analyze the free energy of the monoclinic $P2_1/m$ structure in comparison to the previously assigned orthorhombic $Pnma$ group and find the $P2_1/m$ phase is 0.01 eV/f.u. lower in energy, supporting our single-crystal X-ray diffraction results that the monoclinic distortion is present at ambient conditions. In Figure 3b, the total energy difference between the $P2_1/m$ and $P2_1/c$ model narrows with pressure, exhibiting an energy change of -0.08 eV/f.u. as pressure increases from ambient to ~ 2 GPa. With further compression, a second transition occurs between 4.7 and 5.6 GPa, as illustrated in Figure 3c, showing the unit cell volume evolution up to 10.6 GPa.

In Figure 3a, the significant variation in unit cell volume and calculated energy change between the low- and high-pressure phase at 1.3 GPa coincides with a first-order phase transition, consistent with the reported metastable high-pressure phase produced by sudden pressure release[18]. In addition, the absence of a group-subgroup relationship between $P2_1/m$ and $P2_1/c$ symmetry indicates that the symmetry transformation cannot be ascribed to continuous tuning parameters, such as rotations, displacements, and tilts. Despite $P2_1/c$ (#14) displaying higher symmetry than $P2_1/m$ (#11), the local structural symmetry is decreased within the same unit cell volume. As illustrated in Figure 3d-e, the increased unit cell volume of the $P2_1/c$ structure accommodates the distorted PbBr$_x$ units, disrupting the A-$\bar{\text{A}}$-A-$\bar{\text{A}}$ stacking sequence of the PbBr$_6$ polyhedral layers into an A-B-$\bar{\text{A}}$-$\bar{\text{B}}$ arrangement, owing to the formation of overcoordinated PbBr$_7$ decahedra.

The disproportionate polyhedra deteriorate the long-range order of the inorganic framework, significantly impacting electron transport in this material. In Figure 4a, the $P2_1/m$ lead framework exhibits three-dimensional ordering, evidenced by three distinct {200} peaks corresponding to the Pb interlayer distances along different crystallographic axes (for $P2_1/m$, $a \neq b \neq c$); conversely, this interlayer ordering becomes unidirectional under $P2_1/c$ symmetry, marked by a single (013) peak that represents the scattering of ordered Pb layers. For clarity, the (002) plane in the $P2_1/m$ structure is equivalent to the (013) plane defined in the $P2_1/c$ unit cell, alongside the schematic illustration of this plane shown in Figure 4b. Above 1.3 GPa, significant displacement of Pb atoms leads to the formation of [Pb$_5$Br$_{24}$]$^{14-}$ clusters through the fusion of neighboring PbBr$_6$ units, consisting of edge-sharing PbBr$_6$ octahedral and PbBr$_7$ decahedral, as shown in Figure 4b-c. These clusters act as basic building blocks of the high-pressure structure sandwiched by ordered [PbBr$_6$]$^{4-}$ layers, enhancing the long-range symmetry within an enlarged unit cell.

To examine the impact of these pressure-induced distortions on electron transport properties, we conducted Heyd-Scuseria-Ernzerhof hybrid functional-based DFT calculations to determine the changes in band gap and resistivity under pressure. Figure 5a displays the calculated band gap values of the lowest energy phase at each pressure. At 0.2 GPa, the band gap value of the $P2_1/m$ phase is found to be 2.32 eV, aligning with the experimental value of 2.34 eV at ambient conditions,[48] and it slightly decreases to 2.25 eV as pressure increases to 1.26 GPa. Upon transition to the $P2_1/c$ phase, the band gap is increased by 410 meV to 2.66 eV, then progressively decreases with further pressure increase. The widened band gap leads to higher resistivity, thereby impairing electron transport. As shown in Figure 5b, the relative resistivity ($\rho/\rho_0$) in response to pressure exhibits negligible changes up until the point of phase transition. The deformation of the 3D inorganic framework results in a substantial increase in resistivity by ~ 700 times. Although resistivity gradually recovers with further increased pressure, it remains 100 to 200 times higher than that of the $P2_1/m$ phase. This underscores the critical need for strategies to boost the mechanical strength of the CsPbX$_3$ compounds or adopt alternative protective engineering to mitigate the lattice deformation under mechanical pressure in practical applications.

## 3. Discussion

### 3.1 Comparison with Oxide Systems

Fitting the *P-V* data collected at ambient temperature to a Murnaghan equation of state gives the bulk modulus value of CsPbBr$_3$ to be 6 ±2 GPa, significantly lower than those observed in transition metal perovskite oxide systems RMnO$_3$ (R=La, Dy, Ho, and Lu) with values exceeding 100 GPa.[46-47] Moreover, previous studies using the Murnaghan-Birch equation of state revealed that the bulk moduli of CsPbBr$_3$, MAPbX$_3$, and FAPbX$_3$ are nearly 5 times lower than perovskite titanate oxides, such as CaTiO$_3$, SrTiO$_3$, and BaTiO$_3$.[49] The characteristic difference between lead-based photovoltaic perovskites and oxide perovskites lies in the softness of Pb-Br bonds that deform under low pressure, in contrast to the rigid Ti-O bonds that reinforce TiO$_6$ polyhedral against pressure. The softness of these Pb-X bonds can be improved through cation engineering, where the size and geometry of the A cation interplay with the cuboctahedral cage, thereby modifying the arrangement of the surrounding [PbX$_6$]$^{4-}$ octahedra to stabilize the inorganic framework.

### 3.2 Atomic Motions in CsPbBr$_3$ Under Pressure

Based on the atomic displacement parameters (ADPs) for the Pb, Br, and Cs sites derived from temperature-dependent analyses of crushed single-crystal samples ( Figure S12), we modeled the atomic displacement using a classic potential for these three types of sites. In Figure 5c, the parameters $\alpha$ and $\gamma$ are depicted alongside the extracted potential wells, representing the force constant and quartic anharmonic contribution, respectively. At the potential of $k_B T = 300$ K, Cs atoms display the most extensive motions with an amplitude of 0.25 Å, whereas Br and Pb atoms exhibit lower excursions from the equilibrium positions. The feature of ambient atomic motions lays the groundwork for understanding the impact of external forces, especially pressure, on the stability at individual atomic sites.

The structural stability of halide perovskites largely depends on the size and structural compatibility of A cations interacting with the surrounding $[PbX_6]^{4-}$ octahedra in the crystal lattice. The effect of pressure, as shown in Figure 5d, significantly impacts the bonding environment of the A cations. At 0.2 GPa, the Cs atom is off-centered and accommodated by a $Br_{12}$ cage. As pressure increases to 1.6 GPa, the distribution of Br atoms around the Cs site becomes more ordered, evidenced by the more symmetric first shell near 4 Å, and confined in a $Br_9$ cage with three Br atoms shift away from the Cs atoms to form $[PbBr_7]^{5-}$ decahedral. In contrast, the $PbBr_x$ polyhedra distort with increased pressure, enabled by the reduction of the Cs-Br bond length. These pressure-induced atomic rearrangements can be quantified through the change in Voronoi volume (or Wigner-Seitz cell), given in Figure 5e. The Pb volume slightly increases with pressure, while that of the Cs and Br decreases with pressure. Particularly, the most extensive volume reduction occurs in the A cation at the Cs site, initiating at a lower pressure before the phase transition than the Br and Pb sites. The complex lattice dynamics in response to pressure can be understood by examining the free-energy landscape in the following discussion.

**3.3 System Thermodynamics**

In lead halide perovskite photovoltaics, the A cations ($FA^+$, $MA^+$, and $Cs^+$) exhibit weak interaction with the $PbX_6$ octahedra, particularly for inorganic cations forming secondary hydrogen bonds with energies below 0.1 eV.[50] The instability of the Pb-X framework compared to the oxide perovskites can be elucidated through the relative magnitudes of $-T\Delta S$ and $P\Delta V$ terms in Gibbs free energy. As pressure increases approaching a phase transition, the free energy change between the locally disordered (LD) and locally ordered (LO) phases can be written as $G_{LD} - G_{LO} = U_{LD} - U_{LO} + T(S_{LD} - S_{LO}) + P(V_{LD} - V_{LO})$. Having defined the change, this expression can be simplified into the well-known form as $\Delta G = \Delta U - T\Delta S + P\Delta V$. For the $APbX_3$ materials, increasing pressure reduces the local order of $PbX_6$ polyhedra, resulting in a negative ($-T\Delta S$) term. Unlike oxide systems, these materials are soft, and hence, the change in internal energy ($\Delta U$) does not dominate. In the case of $CsPbBr_3$, increasing the pressure by 2 GPa results in a total energy change of 0.08 eV/f.u., revealing a notably smaller internal energy change across the transition region. On the other hand, the $P\Delta V$ term is negative, approximately 0.08 eV/f.u. Therefore, the pressure-induced phase transition of this material is primarily dominated by the $-T\Delta S$ and $P\Delta V$ terms, coupled with a minor change in internal energy. The pressure-induced deformation of PbBr$_6$ polyhedra reduces local order to accommodate space filling in the lattice. This observation contrasts with the systems composed of rigid basic building units (i.e., atomic groups such as polyhedral), where applying pressure induces a transition to higher symmetry, as seen in the transformation of graphite to diamond under pressure.

The behavior of APbX$_3$ materials under decompression reveals further complexities in their phase and thermodynamic behavior. We note that on the release of pressure from ≈ 10 GPa, the group $P2_1/m$ is recovered. However, the atomic displacement parameters are significantly larger than before the pressure increase (compare Tables S5 and S1). Experiments measuring the structural changes as a function of time after pressure release are planned to fully understand the time dependence of the disordered state or determine if it is static on a device relevant time scale.

The low pressures causing the transitions to the locally disordered phase (local order-disorder transition) can be easily achieved by sample strain or external perturbation.[3, 51] It is possible to trap these materials in the locally disordered $P2_1/c$ phase since transition near 1.3 GPa is first order. These results point to the need to explore methods to enhance the first term ($\Delta U$) in the free energy to stabilize the highly ordered ambient pressure phase.

It is known that structural instability can be suppressed by taking advantage of high entropy ($\Delta S$) to destabilize phase transitions.[52-54] This high entropy is expressed by doping chemical sites (configurational entropy), and/or exploring systems with translational, vibrational, and rotational entropy. In ABX$_3$ systems, the X site can not be randomly doped to create this high entropy configuration since exposure to light induces chemical segregation in systems such as AB(Br/I)$_3$.[55] Therefore, exploring high entropy equivalents via A and B site substitution using atoms of similar size is essential. Machine learning methods combined with DFT simulations to explore the complex A/B site spaces (A= K, Cs, Rb… and B=Pb, Bi, Sn …) were reported to find the optimal stable materials.[56]

**3.4 DFT Analysis of Pressure-Induced Structural Changes**

We performed a detailed analysis of the crystal structures of the (DFT optimized) $P2_1/m$ and $P2_1/c$ phases under increasing pressure. Computing the effective coordination number of the polyhedra in each phase reveals further insights. First, we find that the average Pb-Br bond length is shorter in the $P2_1/m$ phase than the $P2_1/c$ phase, and the bonds compress approximately the same amount versus pressure (Figure S14a). However, under increasing pressure, there is a large bond strain on the Pb-Br bonds in the $P2_1/m$ phase, increasing to up to 4% relative to the low-pressure case at 4.65 GPa. To relieve this, some of the identical PbBr$_6$ octahedra present in $P2_1/c$ disproportionate to 5.5-fold and 6.5-fold distorted octahedra in the $P2_1/c$ phase (the rest remain 6-fold coordinated), reducing the bond strain to less than 3%. This is accomplished via displacements of certain Br atoms, which breaks the corner connectivity of the 6-fold octahedra, and instead forms disconnected layers of overcoordinated and undercoordinated PbBr$_6$ distorted octahedra (which are edge sharing within each layer). Along the *a*-axis of the overcoordinated layers (average effective coordination number of 6.33) alternate with the undercoordinated layers (average coordination number of 5.75). This is summarized in Figure S15.

The changes are even more dramatic in the case of the CsBr$_x$ polyhedra. While these polyhedra are nominally considered to be AX$_{12}$ cuboctahedra in ideal ABX$_3$ perovskites, they are typically undercoordinated. In contrast to the Pb-Br bonds, which are equally compressible in both phases of CsPbBr$_3$, the Cs-Br bonds are more resistant to compression in the $P2_1/c$ phase than $P2_1/m$; furthermore, in agreement with the analysis shown in Figure 5c, the bonds distort more than the Pb-Br bonds. If the average Cs-Br bond length is monitored as a function of pressure (Figure S14b), we find that there is a crossover at 1.26 GPa, exactly the pressure at which the $P2_1/m$ phase transitions to the $P2_1/c$. To understand why, we consider the case of both the $P2_1/m$ phase and $P2_1/c$ phase at the maximum of 4.65 GPa. Under these conditions, the CsBr polyhedra in the $P2_1/m$ phase are under considerable bond strain of 8.5% relative to the 0 pressure case. In the $P2_1/c$ phase, this Cs-Br bond strain is reduced significantly (nearly halved to 4.5%) by an increase in the CsBr$_x$ coordination; this is due to the average coordination number *x* of the CsBr$_x$ polyhedra increasing from 7.7 in $P2_1/m$ to 8.6 in $P2_1/c$. In the $P2_1/c$ phase, there are then layers consisting of half 8.8-fold and half 8.4-fold coordinated CsBr$_x$ polyhedral, which alternate along the *c* axis via a shift of half of a unit cell. This is depicted in Figure S16.

Finally, to better understand the chemical driving force for this rearrangement and polyhedral disproportionation, we computed the partial density of states (pDOS) of each phase under pressure (Figure S17 for $P2_1/m$ and Figure S18 for $P2_1/c$), as well as the Bader charge on each atom in the $P2_1/m$ phase (Figure S19). As pressure increases, the charge on the Cs (Figure S19a) and Pb (Figure S19b) atoms decreases, while those on the Br (Figure S19c) atoms increases (indicative of increased overlap between Cs / Pb and Br), as expected. However, while the charge on all Cs and Pb atoms decreases uniformly, the charge on the Br atoms does not, with half of the Br atoms having a more negative charge than the other half (this split starts to appear close to the phase transition pressure of 1.26 GPa, Figure S19c). We highlighted those Br atoms with the more negative charge in red in Figure S19d. While this change is small, it can provide some indication of the driving force for the phase transition. Furthermore, the increased overlap is reflected in the pDOS. Under pressure, the pDOS of the $P2_1/m$ phases (Figure S17) shows a broadening of the Cs p band between -7 and -8 eV, and a corresponding increase in the overlap of the Br p states and Cs p states. The pDOS of the $P2_1/c$ phases (Figure S18) shows a further increased overlap between the Pb states, Cs p states, and Br p states compared to the $P2_1/m$ phases.

Collectively, as the pressure on the $P2_1/m$ phase increases, large strains are induced on both the Pb-Br and Cs-Br bonds, while the Br atoms are in two different environments. Across the phase transition from $P2_1/m$ to $P2_1/c$ there is a synergistic effect between the Br atoms rearranging so as to relieve the Cs-Br bond strain, in turn breaking the corner connectivity of the PbBr$_6$ octahedra, causing them to be disproportionate (which also relieves the Pb-Br bond strain). As mentioned previously, the Cs-Br polyhedral remains relatively symmetric, whereas the Pb-Br polyhedral is disproportionate in several distorted forms.

## 4. Conclusion and Perspectives

Through synchrotron single crystal X-ray diffraction and DFT calculations, we explore the intrinsic instability of CsPbX$_3$ photovoltaic perovskites, particularly under hydrostatic pressure, detailing atomic fluctuations and structural transitions that affect their performance in devices. Comprehensive reciprocal space analysis reveals the underlying lattice distortions present in ambient $P2_1/m$ structure and further evolves to a distorted $P2_1/c$ phase as pressure increases above 1.3 GPa. This transition, attributed to the inherent lattice softness and weak Pb-X bonds, highlights the intrinsic instability of high-symmetry phases in these materials under stress, corroborated by DFT simulations indicating significant shifts in Gibbs free energy primarily through the pressure-volume ($P\Delta V$) term. We demonstrate the mechanism of symmetry change in long-range order through the local distortion at individual atomic sites, showing a contrary behavior observed in classical oxide perovskites. Our findings not only provide detailed insights into the pressure-induced behaviors of halide perovskites at the atomic level but also underscore the urgent need for material engineering strategies to address the intrinsic instability for advancing perovskite-based technologies.

Analyzing the intrinsic instability of a model ABX$_3$ material uncovers a synergistic effect of the halide atom rearrangement between A and B cations upon pressure. The challenge and opportunity lie in strategically mitigating the pressure-induced bond strains owing to the reduced volume, notably on A-X and B-X bonds, potentially through the incorporation of oversized and undersized A and B cations. This approach could counteract lattice strain more effectively than solely strengthening A-X or B-X bonds. Moreover, the low-pressure-induced first-order phase transition stems from the sudden breaking of Cs-Br bonds, forming overcoordinated PbBr$_x$ polyhedra, which cannot be ascribed to any continuous structural parameters. As such, exploring a broader spectrum of A and B cation stoichiometries, sizes and geometries to maintain the coordinates of the BX$_6$ polyhedra upon external perturbation is the key to mitigating the sudden lattice deformation towards a second-order transition. Addressing this first-order behavior could enhance structural reversibility to recover the performance from mechanical damage in practical use, especially for f-PSCs. Ultimately, it necessitates a deeper understanding of the interrelations between different A cations and the halide perovskite framework, aiming to forge a new frontier in developing perovskite-based devices that exhibit enhanced robustness and operational lifetime.

**Methods**

*Sample Synthesis*: A solution growth method was utilized for crystal preparation,[57] starting from high-purity CsBr and PbBr$_2$ precursors. Orange single crystals were obtained. More details on the methods as well as crystal structure information (experimental data and DFT results) under pressure, are provided in the supplementary information document.

*Single Crystal X-ray Diffraction:* High-pressure single crystal diffraction measurements were performed at APS beamline 13-ID-D and 13-BM-D (GESCARS) at Argonne National Laboratory. The beam size used was 2μm (V) x 3μm (H) at 13-ID-D and 4 μm (V) x 10 μm (H) at 13-BM-D with a wavelength of 0.2952 Å in both cases. The detector-to-sample distance was 233.45 mm at 13-BM-D and 207.07 mm at 13-ID-D (based on a NIST LaB$_6$ standard). A high dynamic range PILATUS 1M detector (maximum count rate = $10^7$ cps/pixel, counter depth =20 bit) with 172 μm x 172 μm CdTe-based pixels, was used. Mini-BX80 diamond anvil cells with 40° two-theta opening (80°full opening), and conical Boehler-Almax design seats and diamond anvils with 400-micron culets were utilized in all measurements. Rhenium gaskets with 200 μm thickness were pre-indented to ~67 μm and 200 μm holes were drilled to create the sample chambers. Neon was gas loaded using the GSECARS/COMPRES gas loading system available at Sector 13, ANL, and served as the pressure-transmitting medium. Ruby balls and Au foils were placed near the samples. Ruby balls were used for the initial calibration and Ne gas loading. Samples were loaded under pressure ≈12 hrs before measurements were conducted. This loading of fresh samples initially held in a moisture free environment was conducted to minimize the exposure of the sample to air and avoid possible chemical alterations. The full range of pressure was calibrated using Au standard diffraction pattern taken at each pressure (SI section for details). Small pressure steps were enabled with the use of a gas membrane apparatus. The ambient pressure pattern was recovered on pressure release. Dioptas[58] was utilized to integrate the two-dimensional diffraction images (powder rings) to generate the intensity vs. 2θ curves used for the pressure calibration (Au powder pattern). LeBail fits of the Au powder patterns were made at each pressure to extract the lattice parameters (SI section for details).

*Powder Diffraction (PDF Refinement)*: Pair distribution function (PDF) measurements were conducted between 10 to 200 K at NSLS-II beamline 28-ID-2 (XPD) beamline at Brookhaven National Laboratory using a wavelength 0.1872 Å. Measurements utilized Perkin Elmer Area detectors with a sample-to-detector distance of 196.6 mm. The detector-to-sample distances were derived by fits to Ni powder calibration standards. The Ni standard was also used to determine the instrumental parameters ($Q_{damp}$ and $Q_{broad}$), which were held fixed for these samples. The Q-range range between $Q_{mim}$ = 1.2 Å$^{-1}$ and $Q_{max}$ = 22.5 Å$^{-1}$ used.

*Density Functional Theory Simulations:* Density functional calculations in the projector augmented wave approach was carried out utilizing the VASP code[59]. Full structural optimization was conducted for both lattice parameters and atomic positions, starting from the experimentally derived unit cells and atomic positions. The LDA exchange functional (Ceperly and Alder as parameterized by Perdew and Zunger[60]) was used to obtain the relaxed structure. The ground-state structure was optimized so that forces on each atom were below 2 x 10$^{-5}$ eV/Å. A 5 × 5 × 5 Monkhorst-Pack[61] k-mesh for the *P2$_1$/m* phase and 1 × 3 × 1 k-mesh for the *P2$_1$/c* phase was used, as well as an 800 eV plane wave cutoff for both. The atom-resolved densities of states were computed using the Perdew-Burke-Ernzerhof (PBE) functional[62]. The band gaps were computed using Heyd-Scuseria-Ernzerhof (HSE) hybrid functional[63] with 20% exact exchange included. The effective coordination number (ECN) of the polyhedral was determined using the VESTA software[64], which computes this value for a given polyhedral *via*:

$$ECN = \sum_i \exp\left(1 - \left(\frac{L_i}{L_{avg}}\right)\right)^6,$$

where $L_i$ is the bond length between the central atom and ligand *i* and $L_{avg}$ is a weighted average of the bond lengths. The resistivity ratio $\rho/\rho_0$ was calculated as:

$$\frac{\rho}{\rho_0} = \frac{\exp(-E_g/kT)}{\exp(-E_{g0}/kT)},$$

where $E_g$ and $E_{g0}$ are the band gaps at the current pressure and 0.2 GPa pressure, respectively, k is the Boltzmann constant, and T is the temperature (298 K). The Bader charge was calculated using the approach of Henklemann *et al*.[65]

*Data Availability:* All data related to this paper and its findings are available from the corresponding authors upon reasonable request.

**Acknowledgments**

This work is supported by NSF Award DMR-2313456. High-pressure Single crystal X-ray diffraction measurements were conducted at beamline 13-ID-D at GeoSoilEnviroCARS (GSECARS, The University of Chicago, Sector 13), Advanced Photon Source (APS), Argonne National Laboratory. GSECARS is supported by the National Science Foundation – Earth Sciences (EAR – 1634415). This research used resources of the Advanced Photon Source, a U.S. Department of Energy (DOE) Office of Science User Facility operated for the DOE Office of Science by Argonne National Laboratory under Contract No. DE-AC02-06CH11357. This research used 28-ID-2 of the National Synchrotron Light Source II, the National Synchrotron Light Source II, a U.S. Department of Energy (DOE) Office of Science User Facility operated for the DOE Office of Science by Brookhaven National Laboratory under Contract No. DE-SC0012704. Single-crystal X-ray diffraction measurements were performed at NSF's ChemMatCARS, Sector 15 at the Advanced Photon Source (APS), Argonne National Laboratory (ANL) which is supported by the Divisions of Chemistry (CHE) and Materials Research (DMR), National Science Foundation, under grant number NSF/CHE- 1834750. This research used resources of the National Energy Research Scientific Computing Center (NERSC), a U.S. Department of Energy Office of Science User Facility located at Lawrence Berkeley National Laboratory, operated under Contract No. DE-AC02-05CH11231. Y. Yan acknowledges support from NSF award 1851747 for perovskite materials synthesis. DFT calculations performed by J. Young were done on the New Jersey Institute of Technology supercomputing cluster (Lochness) and the Carbon cluster at the Center for Nanoscale Materials (Argonne National Laboratory) under allocations CNM72794, CNM75950, and CNM82958. Work performed at the Center for Nanoscale Materials, a U.S. Department of Energy Office of Science User Facility, was supported by the U.S. DOE, Office of Basic Energy Sciences, under Contract No. DE-AC02-06CH11357. We are indebted to C. L. Dias of the NJIT Physics Department for constructive criticism and physical insight, which has significantly improved the manuscript.

**Author Contributions**

High-pressure X-ray diffraction measurements and analysis were conducted by S.L., T.A.T., S.C. and V.P. PDF measurements and analysis were conducted by S.L., T.A.T and S.G. DFT modeling was conducted by M.L., J.Y. and T.A.T. Samples used were prepared by Y.Y. The manuscript was prepared by S.L., J.Y. and T.A.T.

**Competing Interest**

The authors declare no competing interest.

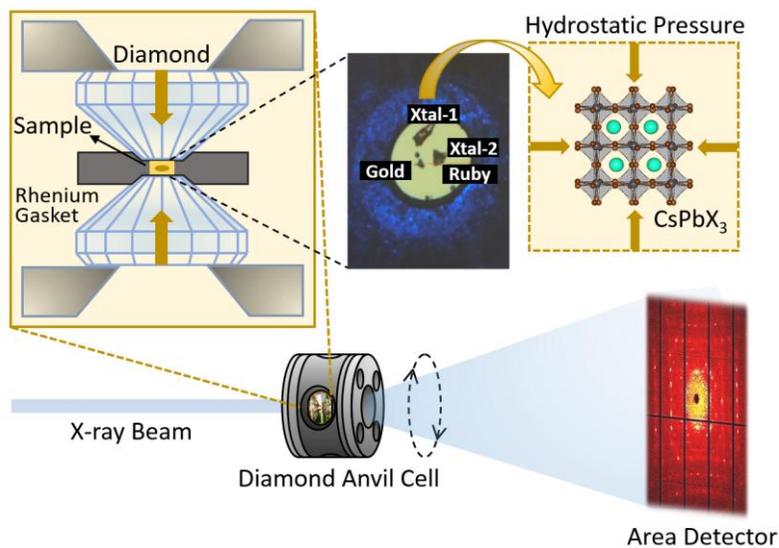

**Figure 1.** Schematic of the experimental setup. A diamond anvil cell with a large opening angle is used to compress the studied crystals. A pixel array area detector is used to collect diffraction patterns. The area detector image for a single angular position is shown. The top middle figure shows the single crystal samples, ruby, and gold, inside the diamond anvil cell. The light (yellow) area is approximately 200 μm in diameter. The single crystals are under hydrostatic pressure in a Ne pressure medium to investigate the structural deformation modes.

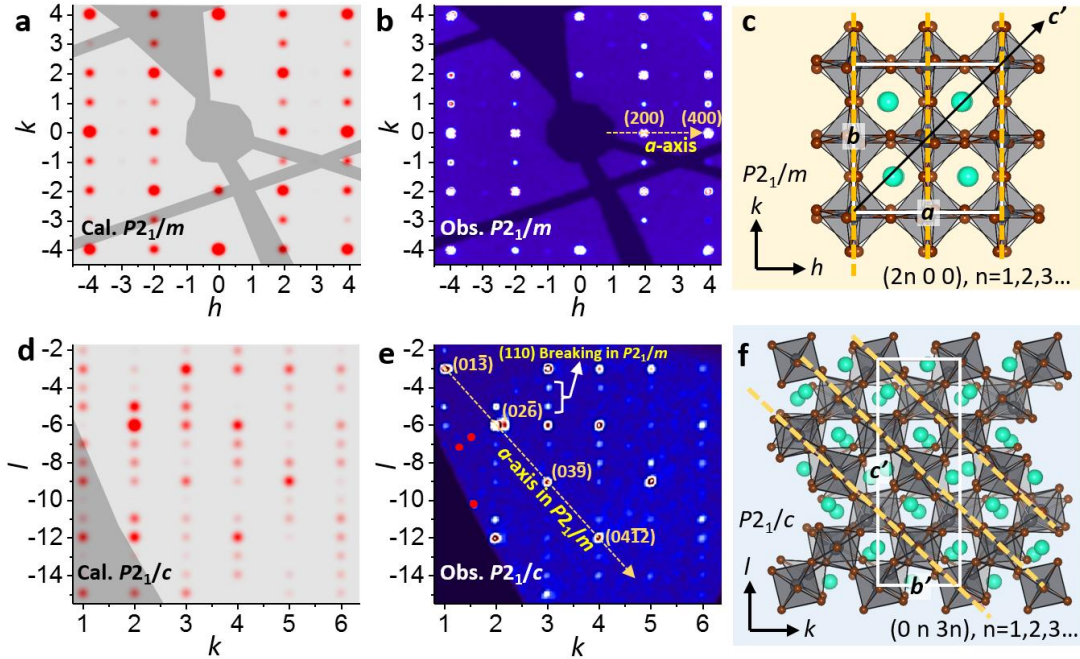

**Figure 2.** Reciprocal lattice and symmetry breaking during phase transition. (a) Calculated reciprocal lattice (*hk0*) plane for P = 0.2 GPa in $P2_1/m$ space group below the structural transition. The inclusion of weak reflections (*h* is odd) reduces the space group symmetry from orthorhombic *Pnma* to the more accurate space group assignment of monoclinic $P2_1/m$. (b) Observed reciprocal lattice (*hk0*) planes at P = 0.2 GPa. The arrow from (200) to (400) reflection indicates the direction of the *a*-axis in the $P2_1/m$ unit cell dimension $2a_0 \times 2a_0 \times 2a_0$, where $a_0$ is the shortest Pb-Pb distance in the inorganic framework. (c) Corresponding $P2_1/m$ structure with (2n 0 0) planes stacking along the *a*-axis. (d) Calculated reciprocal lattice (*hk0*) plane for P = 1.6 GPa in $P2_1/c$ space group above the structural transition, in good agreement with the observed reflections in (e). The arrow indicates the stacking direction of (0 n 3n) planes, which is the *a*-axis direction in the $P2_1/m$ unit cell configuration. The *c*-axis direction (denoted as *c'*) in the $P2_1/c$ unit cell is diagonal of the *ab* plane in $P2_1/m$ cell, along [110] direction. Breaking symmetry along the (110) plane leads to an elongated *c'*, along the diagonal of the original $P2_1/m$ cell configuration with a length of $3\sqrt{2}a_0$. (f) $P2_1/c$ unit cell configuration with (0 n 3n) planes.

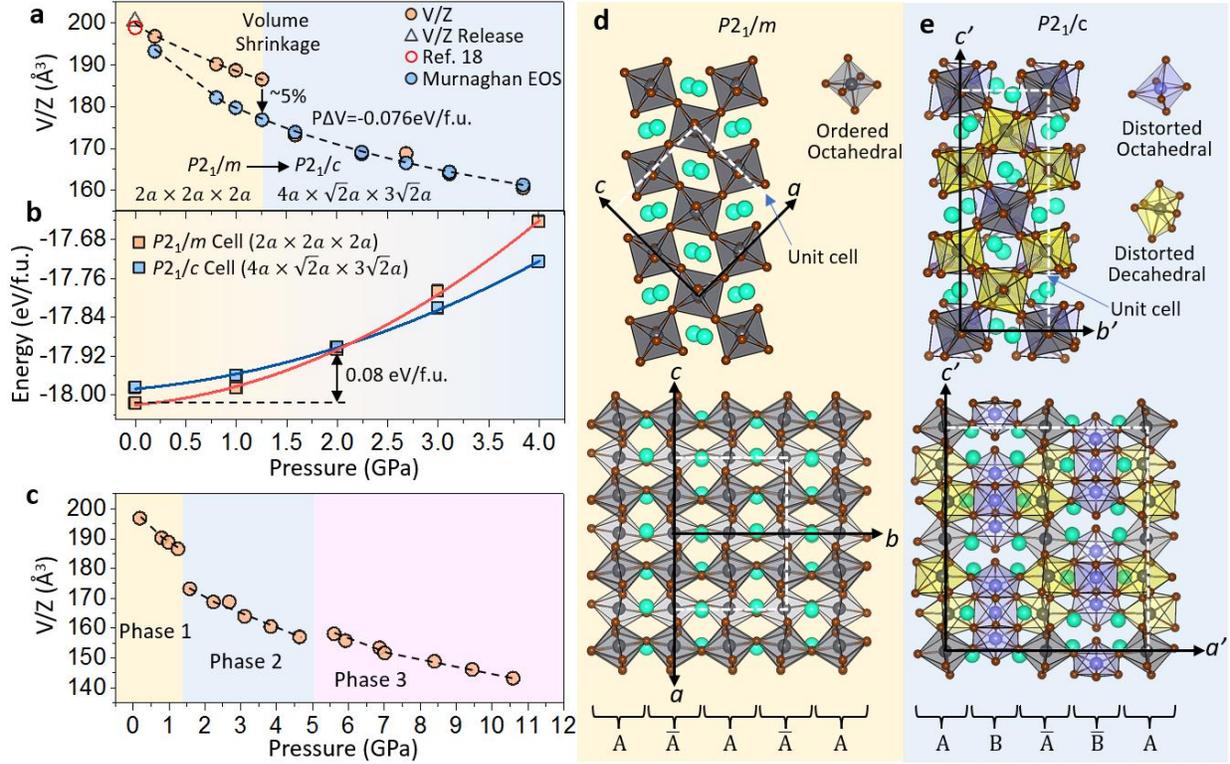

**Figure 3.** Volume contraction dictated internal energy change during pressurization. (a) Unit cell volume per CsPbBr$_3$ molecular unit (V/Z or volume per formula unit (f.u.)) from single-crystal structural solutions under pressures. The V/Z after pressure release and standard single-crystal solutions (ambient) are given as triangle and open-circle symbols, respectively. Crystal structural transitions to lower symmetry space groups were found near 1.3 and 4.7 GPa. The change near 1.3 GPa is abrupt. The Murnaghan equation of state fits the data from 1. 3 GPa to 4.7 GPa, yielding K = 6±2 GPa and K' =9±2 GPa for the bulk modulus and its derivative at ambient pressure, respectively. The extrapolation of this model to low pressure reveals the PΔV value at the transition (~0.076 eV/f.u.) with a 5% volume change. (b) Total energy vs. pressure curves for the ambient pressure phase $P2_1/m$ and the higher-pressure phase $P2_1/c$ (c) V/Z evolution for the full pressure range studies from ambient to 10.6 GPa showing two phase transitions. (d) Schematic of $P2_1/m$ unit cell configuration. PbBr$_6$ polyhedra layers stack along the $b$-axis with an A-$\bar{\text{A}}$-A-$\bar{\text{A}}$ stacking sequence. (e) ) Schematic of $P2_1/c$ unit cell configuration. The PbBr$_6$ polyhedra layer is distorted, showing an A-B-A-B stacking sequence, where A layer is composed of distorted decahedral and B layer is constituted of distorted octahedra.

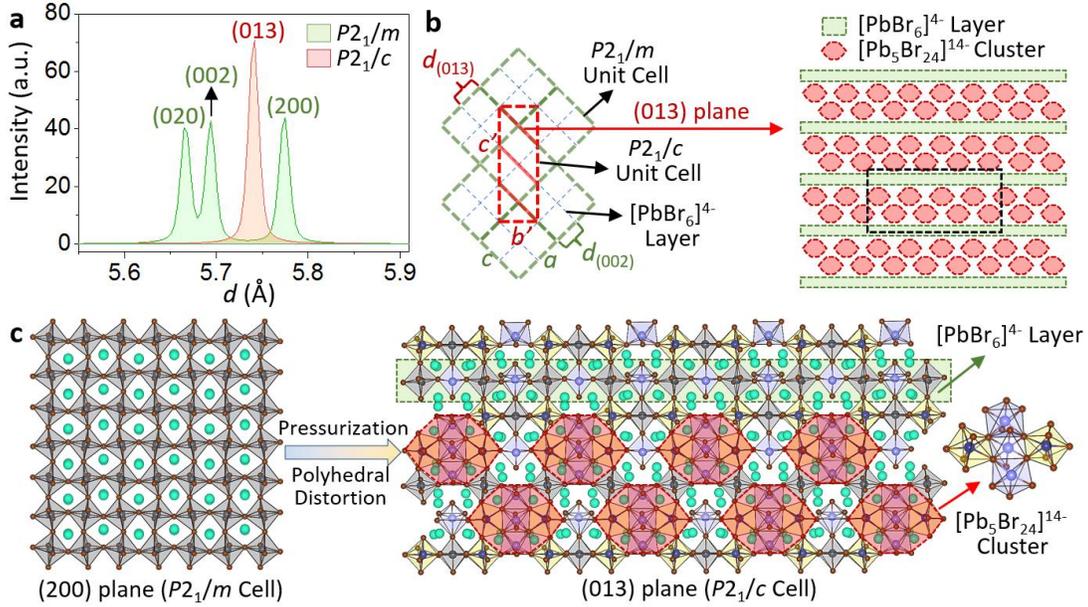

**Figure 4.** Pressure-induced distortions in the inorganic framework. (a) Computed diffraction lines from the PbBr$_6$ polyhedral layers, denoted as {200} planes in $P2_1/m$ symmetry and the (013) plane for $P2_1/c$ symmetry. (b) Geometric illustration of the atomic relationship between $P2_1/m$ and $P2_1/c$ unit cell. The green bold dashed square indicates the $P2_1/m$ unit cell, and the red bold dashed rectangle is the $P2_1/c$ unit cell. The (002) layer in $P2_1/m$ is denoted as (013) under the symmetry of $P2_1/c$. Above 1.3 GPa, polyhedral distortions and lattice rearrangements occur. The formed [Pb$_5$Br$_{24}$]$^{14-}$ clusters (red) sandwiched by [PbBr$_6$]$^{4-}$ polyhedral layer (green) act as basic building units in the high-pressure phase. (c) Pressure-induced lattice distortion and rearrangement correspond to the structure in the black dashed rectangle in (b). The three-dimensional ordered $P2_1/m$ structure evolves to a disordered $P2_1/c$ structure composed of [Pb$_5$Br$_{24}$]$^{14-}$ clusters (red shaded) and [PbBr$_6$]$^{4-}$ polyhedral layers (green shaded). The formation of overcoordinated and undercoordinated PbBr$_x$ polyhedra reduces local symmetry. However, in a longer structural range, the enlarged [Pb$_5$Br$_{24}$]$^{14-}$ clusters (compared to [PbBr$_6$]$^{4-}$ polyhedra) act as basic building units that enhance symmetry within an expanded unit cell.

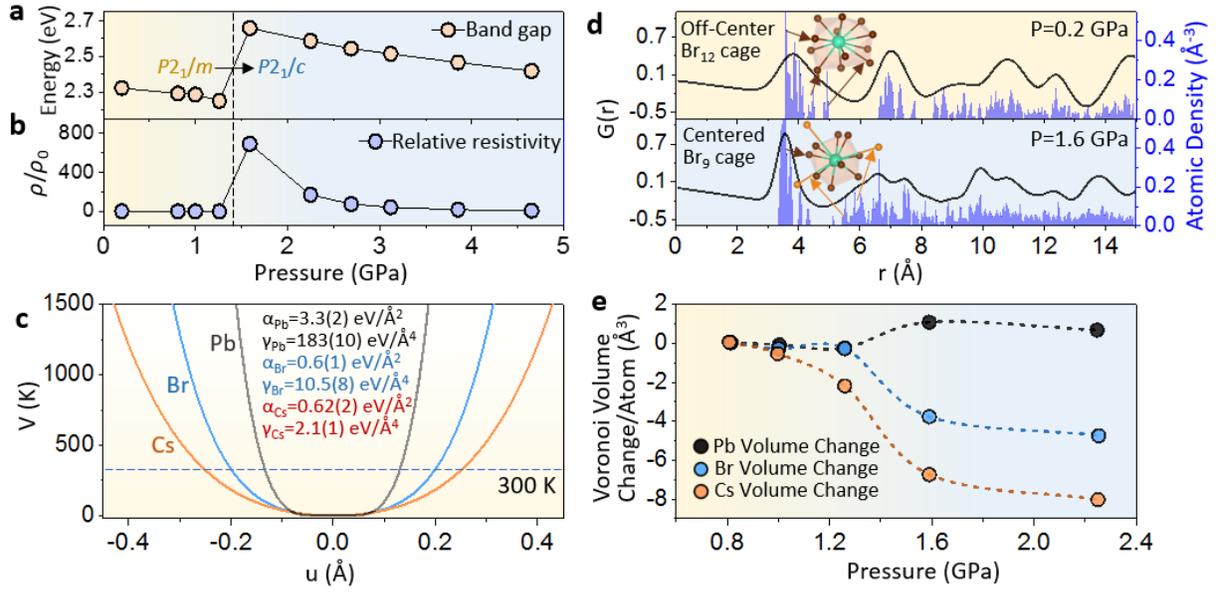

**Figure 5.** Influence of Cs and Br softness on structural stability. (a) Band gap of the lowest energy phase as a function of pressure as computed by HSE DFT. (b) Resistivity of the lowest energy phase at each pressure ($\rho$) normalized to the resistivity of the CsPbBr$_3$ $P2_1/m$ 0.2 GPa phase ($\rho_0$). The dashed line indicates the pressure-induced phase transition from $P2_1/m$ to $P2_1/c$. (c) Potential wells representing the motion of the average Pb, Br, and Cs atoms. (d) Labeled distorted polyhedral in the high-pressure structure (1.6 GPa). Distribution of the Br atoms about Cs in the low-pressure structure at 0.2 GPa and the high-pressure 1.6 GPa structure. The Cs atom is accommodated by the Br$_{12}$ cage at low pressure with Cs off-centered. At 1.6 GPa, Cs atoms are centered and confined in the Br$_9$ cage. Three Br atoms shift away from the Cs atoms, forming [PbBr$_7$]$^{5-}$ decahedral. Note that the distortions of the PbBr$_6$ polyhedra are enabled by the reduction of the Cs-Br bond length (reduction of Cs size). The square-well-like potentials for Cs and Br enable the large distortions of the polyhedra. (e) Computed Voronoi volume vs. pressure for Pb, Br, and Cs.

Supporting Information

**Origin of Pressure-Induced Structural Instability in CsPbX$_3$ Photovoltaic Perovskites**

*Sizhan Liu, Sandun Amarasinghe, Mo Li, Stella Chariton, Vitali Prakapenka, Sanjit K. Ghose, Yong Yan, Joshua Young[*] and Trevor A. Tyson[*]*

# Supporting Information

**Origin of Pressure-Induced Structural Instability in CsPbX$_3$ Photovoltaic Perovskites**

*Sizhan Liu, Sandun Amarasinghe, Mo Li, Stella Chariton, Vitali Prakapenka, Sanjit K. Ghose, Yong Yan, Joshua Young[*] and Trevor A. Tyson[*]*

### A. Sample Synthesis

To obtain CsPbBr$_3$ crystals, a solution growth method was utilized [1], starting from high-purity CsBr and PbBr$_2$ precursors. Orange single crystals were obtained. For powder samples, crystals were crushed and sieved (~500 mesh). All reported measurements are based on crystal-derived material.

### B. Single Crystal X-Ray Diffraction

High-pressure single crystal x-ray diffraction measurements were performed at APS beamline 13-ID-D and 13-BM-D (GESCARS) at Argonne National Laboratory as detailed in the main text. We note that for calibration in addition to LaB6, a single crystal of orthoenstatite (($Mg_{1.93}$,$Fe_{0.06}$)($Si_{1.93}$,$Al_{0.06}$)$O_6$, *Pbca*, $a$ = 18.2391(3), $b$ = 8.8117(2), $c$ = 5.18320(10) Å) was used to calibrate the instrument model of CrysAlisPro (the sample-to-detector distance, the detector's origin, offsets of the goniometer angles, rotation of the X-ray beam and the detector around the instrument axis).

Small pressure steps were enabled with the use of a gas membrane apparatus. The ambient pressure pattern was recovered on pressure release. Dioptas [2] were utilized to integrate the two-dimensional diffraction images (powder rings) to generate the intensity vs. 2θ curves used for the pressure calibration (Au powder pattern). LeBail fits of the Au powder patterns were made at each pressure to extract the lattice parameters. Based on the compression curve for bulk Au, the pressure values were determined from the lattice parameters based on an expression (Eq. 1 below) derived from the compression data of Fei *et al.* [3]. Between ambient pressure and 76 GPa [3], the pressure as a function of lattice parameter can be expressed as (Mathematica$^{TM}$ expression):

```
Pr[a_] :=
 67.08182869964592` +
  1.0812443865191471`*^18 /
   (-4.08102049632277`*^50 + 1.0751358904719677`*^50 a +
      4.6371023074156295`*^21
      √(7.749732769972096`*^57 - 4.08102049632277`*^57 a + 5.375679452359838`*^56 a²) )^(1/3) -
  2.3862177148667255`*^-15
   (-4.08102049632277`*^50 + 1.0751358904719677`*^50 a +
      4.6371023074156295`*^21
      √(7.749732769972096`*^57 - 4.08102049632277`*^57 a + 5.375679452359838`*^56 a²) )^(1/3)
```
(1)

where a is the lattice parameter and the output (Pr) is the pressure.

Data were collected as images by scanning omega from -35° to +35 in steps of 0.5° resulting in 140 images at each pressure for each $CsPbBr_3$ crystal. Data processing was performed with CrysAlis[Pro] [4]. Using the SCALE3 ABSPACK scaling algorithm Unit cell indexing was conducted at all pressures. Pressure-dependent structural solutions were obtained *via* intrinsic phasing methods using ShelXT and refined using ShelXL in the Olex2 graphical user interface [5]. Structural solutions were obtained between ambient pressure and 4.7 GPa. Above 4.7 GPa, the unit cell was found by indexing the diffraction data, but due to the low structural order, no structural solution was found. For phase 1 anisotropic atomic displacement parameters ($U_{ij}$) were determined for each atom but due to the high number of free parameters due to lowering of symmetry and increase in cell size only the anisotropic parameters ($U_{eq}$) were determined in phase 2.

## C. Powder Diffraction (PDF Refinement)

Pair distribution function (PDF) measurements were conducted done 10 to 200 K at NSLS-II beamline 28-ID-2 (XPD) beamline at Brookhaven National Laboratory using a wavelength of 0.1872 Å. Samples were measured in 1 mm Kapton capillaries with 50-micron-thick walls. Scans were collected with blank capillaries to determine the background scattering. This background was subtracted from all datasets (see data reduction methods in Refs. [6]). For the fits in R-space, the range $2.0 < r < 30$ Å was utilized. The time interval between temperature points was ~2 minutes. Combined with the small temperature steps, the approach kept the samples from being in a quenched state. The data sets were taken from the experiments in our work in Ref [7].

The extracted atomic displacement parameters (ADPs) were fit to an effective potential $V(u) = \frac{\alpha}{2}u^2 + \gamma u^4$ (see Refs. [8,9]) with the isotropic ADP given by $U(\text{Å}^2) =$ $\int_0^\infty x^4 \, Exp\left[-\frac{V(x)}{k_B T}\right] dx \, / \int_0^\infty x^2 \, Exp\left[-\frac{V(x)}{k_B T}\right] dx$ assuming that the motion of the atoms follows a Boltzmann distribution.

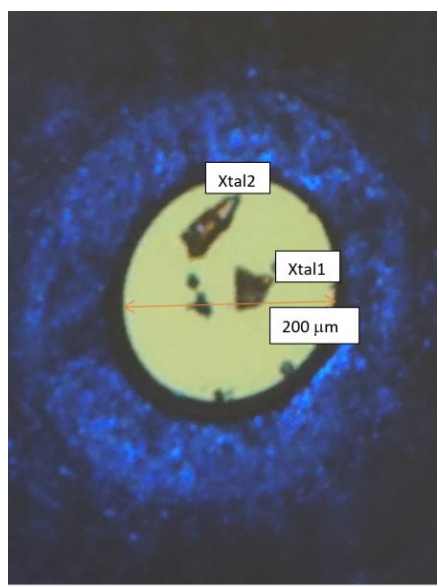

**Figure S1**. CsPbBr$_3$ single crystals in diamond anvil cell. The diamond culet was 400 μm in diameter and the sample hole was 200 μm. The two crystals were examined in high-pressure measurements (~30 μm thick). They were: xtal1 with dimensions ~30 μm x ~25 μm x ~25 μm and xtal2 with dimensions ~30 μm x ~70 μm x ~20 μm. The indented Re gasket was ~67 μm thick. The additional materials on the diamond culets are the ruby chips (spherical shape) and Au metal calibration foils.

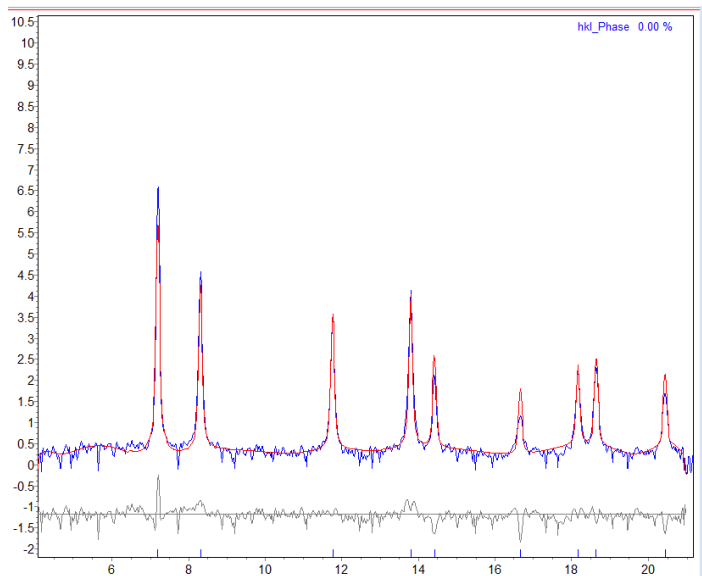

**Figure S2**. Representative LeBail fit showing $I^{1/2}$ vs 2θ for the first Au calibration data set (data =blue, fit=red). Using the extracted lattice parameter yielded a pressure of 0.2 GPa. Structure fits were conducted utilizing Topas software [10].

**Table S1. Pressure List**
(Gold Standard Compression)

**Pressure (Gpa) Au Lattice Parameter (Å)**

| | |
|---|---|
| **Phase 1** | |
| 0.20 | 4.07543 |
| 0.81 | 4.07104 |
| 1.00 | 4.06966 |
| 1.26 | 4.0678 |
| **Phase 2** | |
| 1.59 | 4.0655 |
| 2.25 | 4.06085 |
| 2.69 | 4.05777 |
| 3.12 | 4.05481 |
| 3.85 | 4.04986 |
| 4.65 | 4.04454 |
| **Phase 3** | |
| 5.61 | 4.0382 |
| 5.93 | 4.03609 |
| 6.88 | 4.02996 |
| 7.02 | 4.02911 |
| 8.41 | 4.02036 |
| 9.46 | 4.01391 |
| 10.6 | 4.00701 |
| **Ambient (Pressure Release)** | |

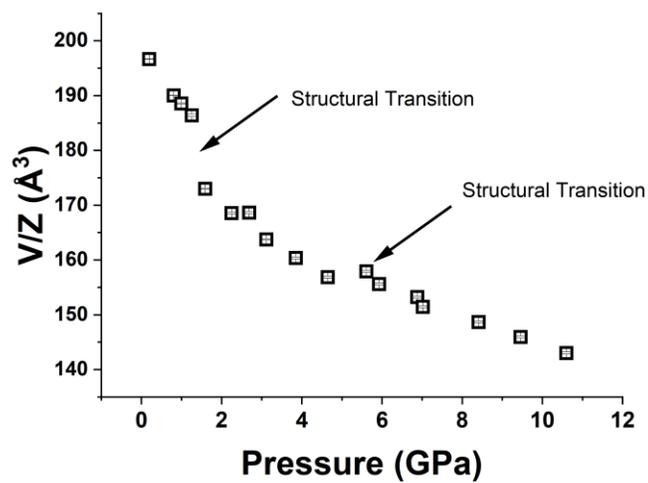

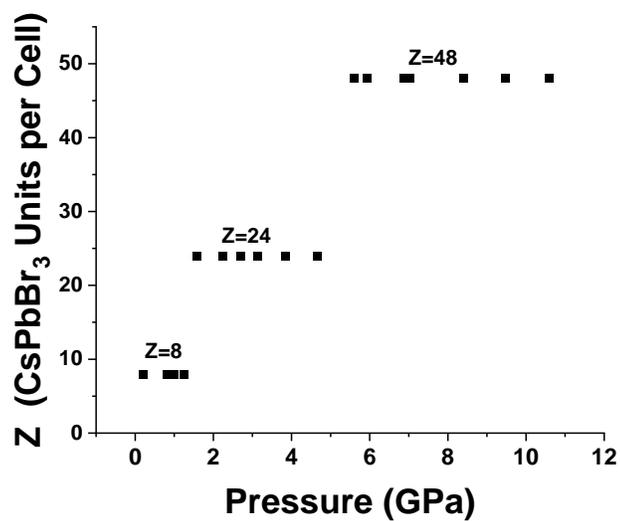

**Figure S3.** (a) Cell volume per CsPbBr$_3$ unit (Z) as a function of pressure. (b) Z for the unit cell as a function of pressure.

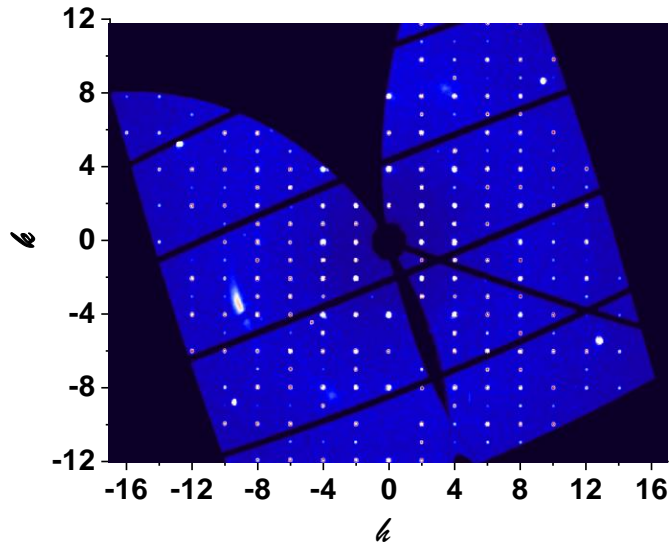

**Figure S4**. CsPbBr$_3$ reciprocal lattice (h0l) at P= 1.3 GPa. For this region of pressure, the ambient pressure cell (~11 Å x ~11 Å x ~11 Å ) persists with pace group *P2$_1$/m*.

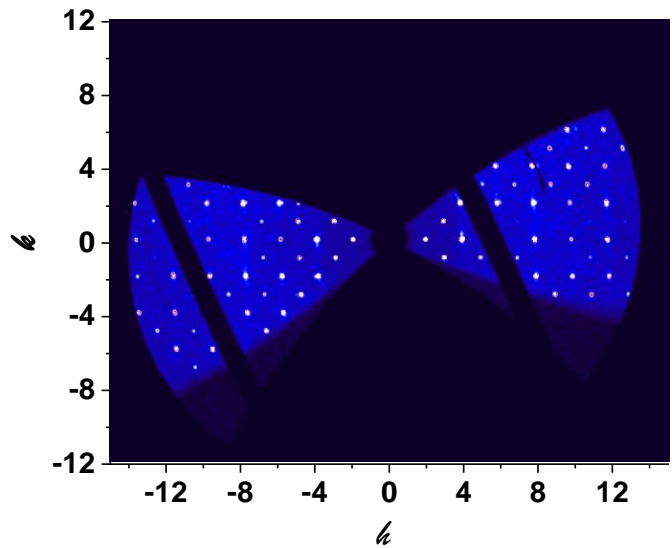

**Figure S5**. CsPbBr$_3$ reciprocal lattice (hk0) plane at P~ 1.3 GPa.

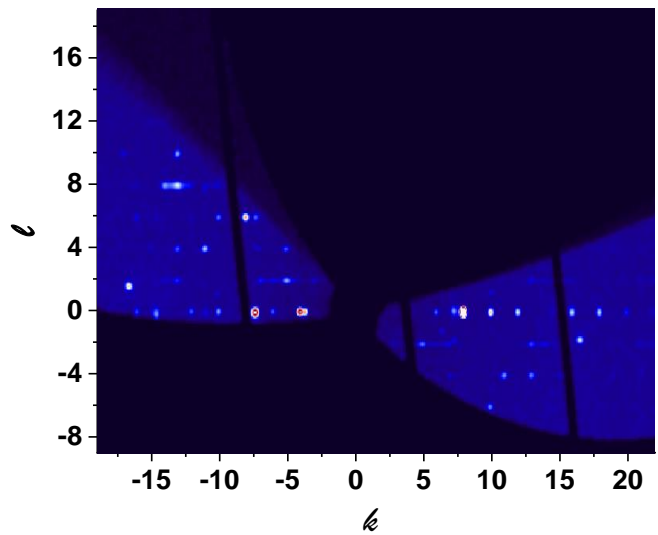

**Figure S6**. Reciprocal lattice (0kl) planes for P = 1.6 GPa, above the first structural transition. Indices are relative to the ~8Å x ~21 Å x ~23 Å unit cell.

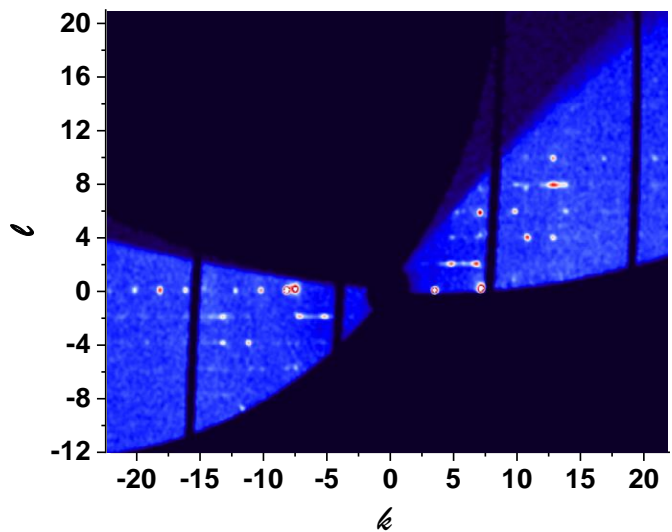

**Figure S7**. Reciprocal lattice (0kl) planes for P = 4.7 GPa, just before the second structural transition. As in Figure S6, indices are relative to the ~8Å x ~21 Å x ~23 Å unit cell.

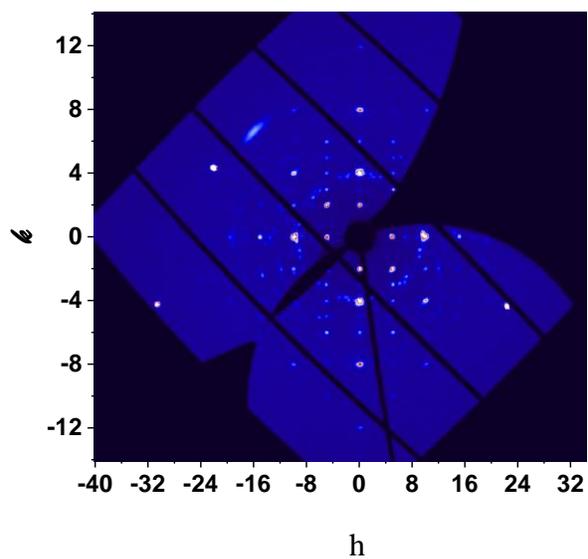

**Figure S8**. Reciprocal lattice (hk0) planes for P = 5.6 GPa, above the second structural transition. Relative to ~11Å x ~26 Å x ~26 Å unit cell.

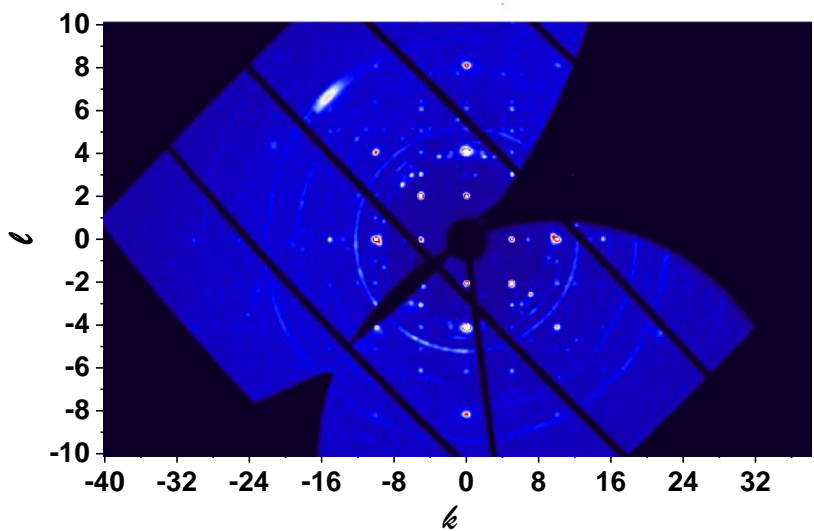

**Figure S9**. Reciprocal lattice (0kl) planes for P = 10.6 GPa, above the second structural transition. Relative to ~11Å x ~26 Å x ~26 Å cell.

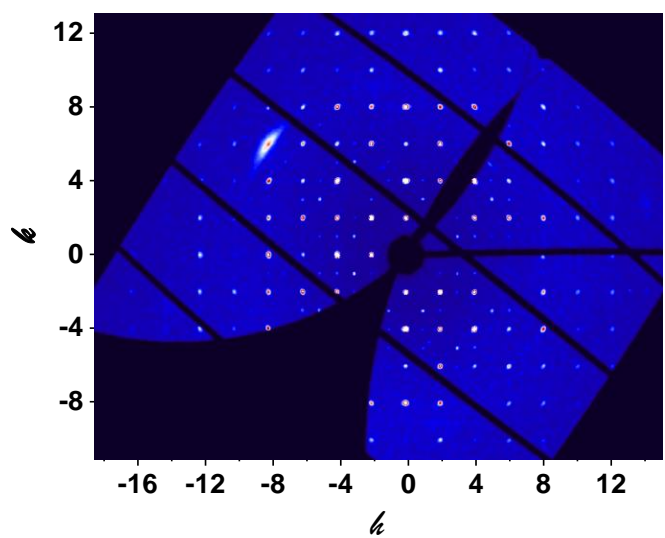

**Figure S10**. Reciprocal lattice (0kl Reciprocal lattice (hk0) planes after releasing the pressure from a maximum value of 10.6 GPa back to ambient pressure.

1 GPa

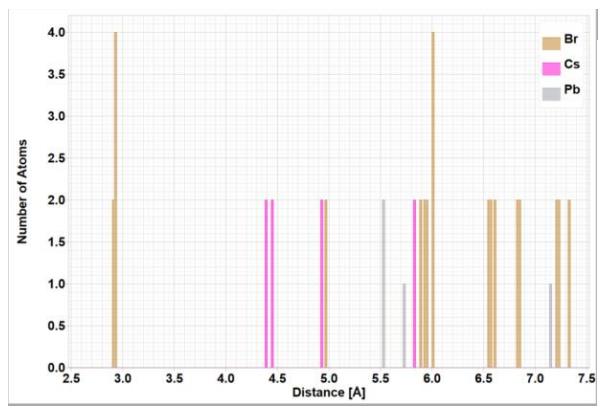
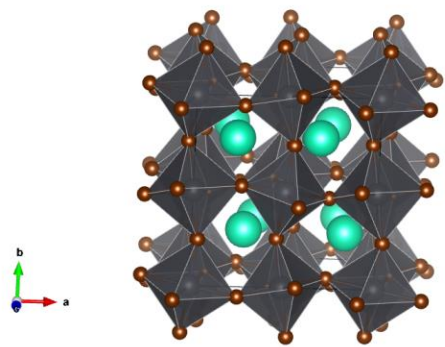

3 GPa

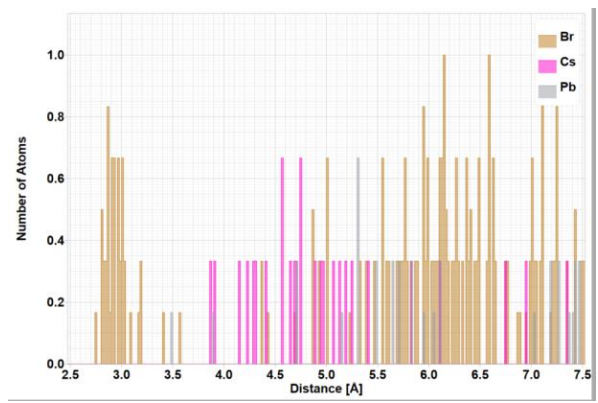
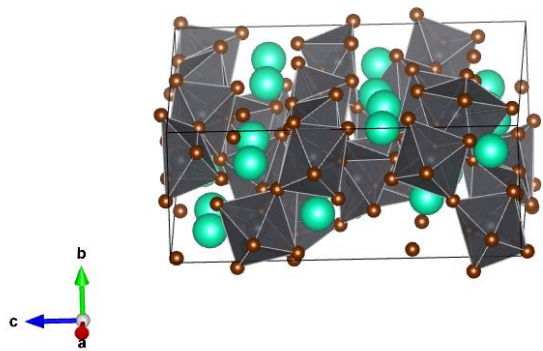

**Figure S11**. Bond distribution about the average Pb site for P ~ 1 GPa (a) and P ~3 GPa. Note that the reduction is in the order occurring on passing through the first structural transition.

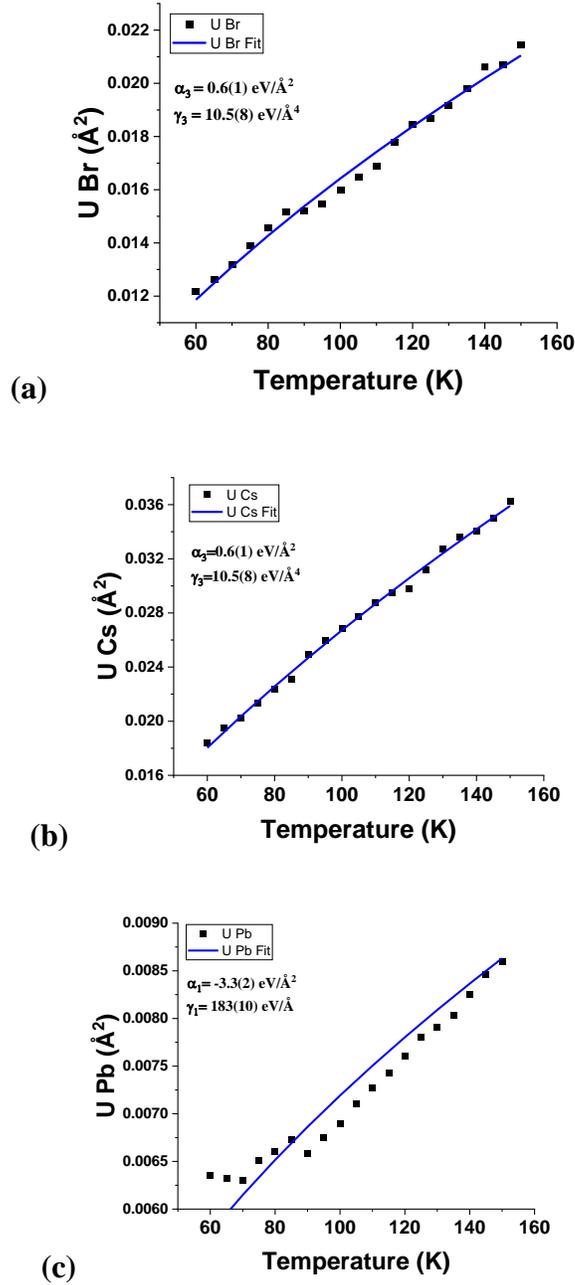

**Figure S12.** Fits for the isotropic atomic displacement parameters U (between 10 and 160 K) for the average Br (a), Cs (b), and Pb (b) atoms to the PDF data using a single particle potentials V(u) assuming a Boltzmann energy distribution. The potential form used was $\frac{\alpha}{2}u^2 + \gamma u^4$ assuming a spherical shape of the potential. The extracted one-particle potentials are given in Figure 5(a) of the main text and Figure S13 below. The points correspond to the experimental data and the solid lines are the atomic model fits.

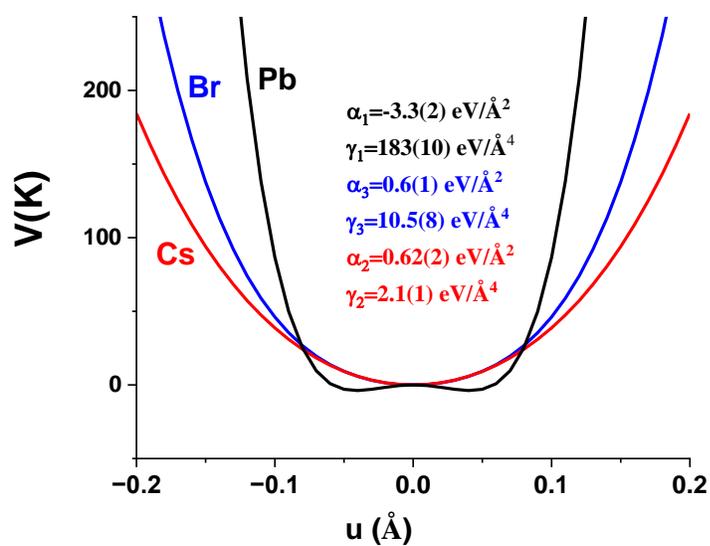

**Figure S13.** Atomic potentials from powder diffraction data based on the temperature-dependent atomic displacement parameters (Figure S12). Scale is expanded compared to that in Figure 5(a).

**Table S2.** Structural Parameters at 0.2 GPa ($P2_1/m$ Space Group, Phase 1)

| Atoms | x(×10⁴) | y(×10⁴) | z(×10⁴) | U$_{eq}$ (Å³×10³) | | |
|---|---|---|---|---|---|---|
| Pb1 | 0 | 5000 | 5000 | 32.6(11) | | |
| Pb2 | 5000 | 5000 | 5000 | 29.9(11) | | |
| Pb3 | 5000 | 5000 | 0 | 35.7(11) | | |
| Pb4 | 0 | 5000 | 0 | 35.7(11) | | |
| Cs1 | 2631(2) | 2500 | 7311(7) | 87(4) | | |
| Cs2 | 2347(2) | 7500 | 7773(7) | 98(4) | | |
| Cs3 | 2740(3) | 2500 | 2297(9) | 93(5) | | |
| Cs4 | 2253(4) | 7500 | 12585(10) | 103(6) | | |
| Br1 | 2496.9(16) | 4730(3) | 4515(6) | 64(3) | | |
| Br2 | 2505.5(15) | 5264(3) | 10445(6) | 80(3) | | |
| Br3 | -239(4) | 7500 | 9728(10) | 80(5) | | |
| Br4 | 5262(5) | 7500 | 10220(10) | 75(5) | | |
| Br5 | 280(5) | 7500 | 5237(11) | 98(6) | | |
| Br6 | 4737(4) | 7500 | 4716(10) | 78(5) | | |
| Br7 | 4526(3) | 5288(3) | 7487(8) | 83(5) | | |
| Br8 | 479(3) | 4740(3) | 7508(8) | 77(5) | | |
| U$_{ij}$ (Pb1) | 26.6(5) | 26.0(4) | 45(3) | 1.0(7) | 0.2(8) | 2.0(3) |
| U$_{ij}$ (Pb2) | 27.0(5) | 29.4(5) | 33(4) | 0.1(7) | -5.7(8) | 0.6(3) |
| U$_{ij}$ (Pb3) | 26.4(5) | 27.3(5) | 53(4) | -0.3(7) | -7.6(8) | 0.7(3) |
| U$_{ij}$ (Pb4) | 25.7(5) | 25.7(4) | 57(4) | -2.9(7) | -2.2(8) | -0.5(3) |
| U$_{ij}$ (Cs1) | 53.8(12) | 57.7(15) | 148(13) | 0 | -21(2) | 0 |
| U$_{ij}$ (Cs2) | 4.8(12) | 58.4(15) | 182(13) | 0 | 6(2) | 0 |
| U$_{ij}$ (Cs3) | 100(2) | 55.3(16) | 124(15) | 0 | 4(3) | 0 |
| U$_{ij}$ (Cs4) | 124(3) | 58.5(18) | 125(19) | 0 | -33(4) | 0 |
| U$_{ij}$ (Br1) | 28.7(10) | 82.8(18) | 81(10) | -6(3) | -5.0(17) | 1.2(8) |
| U$_{ij}$ (Br2) | 22.0(9) | 88.1(19) | 131(10) | -8(3) | 2.1(18) | -0.1(8) |
| U$_{ij}$ (Br3) | 81(2) | 29.0(13) | 131(16) | 0 | -4(4) | 0 |
| U$_{ij}$ (Br4) | 16(3) | 31.6(15) | 77(17) | 0 | -13(5) | 0 |
| U$_{ij}$ (Br5) | 108(3) | 25.6(13) | 161(18) | 0 | -15(5) | 0 |
| U$_{ij}$ (Br6) | 90(3) | 25.5(13) | 119(16) | 0 | -20(4) | 0 |
| U$_{ij}$ (Br7) | 76(2) | 88(2) | 85(17) | 14(3) | 1(3) | 14.5(14) |
| U$_{ij}$ (Br8) | 75(2) | 89(2) | 68(17) | -13(3) | -5(3) | 5.1(14) |

Space Group: $P2_1/m$
$a = 11.588(1)$ Å, $b = 11.722(1)$ Å, $c = 11.58(1)$ Å, $\beta = 90.51(2)°$, Dx = 4.896 g/cm³
$CsPbBr_3$, Z=8
Volume = 1572.90 Å³
Measurement Temperature: 298 K
Crystal Dimensions: ~30 mm x ~25 mm x ~25 mm
Wavelength: 0.29521 Å
2θ range for data collection: 3.254° to 28.224°
Index ranges: $-16 \leq h \leq 17$, $-16 \leq k \leq 18$, $-9 \leq l \leq 6$
Total reflections collected: 3846
Twin law: 1 0 0 0 -1 0 0 0 -1 2

**Table S2. Continued**

Independent reflections:2045  
$R_{int}$ =3.80%, $R_{sigma}$=3.85%  
Number of fitting parameters: 110  
Largest diff. peak/hole: 1.45(Cs3)/-1.64(Pb1) e$^-$/Å$^3$  
$R_1$ = 5.36 %, $wR_2$ = 17.32 %, Goodness of Fit = 1.051

*Atomic displacement parameters $U_{ij}$ (Å$^2$×10$^3$) are in the order $U_{11}$, $U_{22}$, $U_{33}$, $U_{23}$, $U_{13}$, $U_{12}$.

**Table S3.** Structural Parameters at 1.3 GPa (*P*2$_1$/*m* Space Group, Phase 1)

| Atoms | x(×10⁴) | y(×10⁴) | z(×10⁴) | U$_{eq}$ (Å³×10³) | | |
|---|---|---|---|---|---|---|
| Pb1 | 0 | 5000 | 5000 | 25.8(9) | | |
| Pb2 | 5000 | 5000 | 5000 | 28.2(10) | | |
| Pb3 | 5000 | 5000 | 10000 | 28.0(9) | | |
| Pb4 | 0 | 5000 | 10000 | 27.0(10 | | |
| Cs1 | 2713.3(16) | 2500 | 7099(4) | 64(2) | | |
| Cs2 | 2291.4(15) | 7500 | 7918(4) | 62(2) | | |
| Cs3 | 2911.2(19) | 2500 | 2264(4) | 60(3) | | |
| Cs4 | 2094.6(19) | 7500 | 12719(5) | 68(3) | | |
| Br1 | 2478.7(12) | 4684.9(17) | 4416(4) | 51.2(19) | | |
| Br2 | 2518.6(12) | 5314.3(17) | 10577(4) | 55.0(19) | | |
| Br3 | -313(2) | 7500 | 9699(6) | 82(3) | | |
| Br4 | 5300(3) | 7500 | 10316(6) | 64(3) | | |
| Br5 | 289(3) | 7500 | 5305(6) | 78(3) | | |
| Br6 | 4696(2) | 7500 | 4691(6) | 71(3) | | |
| Br7 | 4421.2(18) | 5321(2) | 7486(5) | 64(3) | | |
| Br8 | 580.3(17) | 4681(2) | 7529(4) | 50(2) | | |
| U$_{ij}$ (Pb1) | 30.5(4) | 25.5(4) | 22(3) | 1.3(4) | 2.6(7) | 3.0(2) |
| U$_{ij}$ (Pb2) | 27.7(4) | 24.9(3) | 32(3) | -3.5(4) | 3.6(7) | 0.0(2) |
| U$_{ij}$ (Pb3) | 28.5(4) | 23.6(3) | 32(3) | -1.1(4) | 3.8(7) | 2.6(2) |
| U$_{ij}$ (Pb4) | 29.1(4) | 24.7(3) | 27(3) | -1.5(4) | 3.3(7) | 0.2(2) |
| U$_{ij}$ (Cs1) | 54.8(9) | 48.5(9) | 90(7) | 0 | 2.1(16) | 0 |
| U$_{ij}$ (Cs2) | 52.7(9) | 46.5(9) | 88(7) | 0 | -4.3(15) | 0 |
| U$_{ij}$ (Cs3) | 95.4(15) | 48.8(10) | 34(9) | 0 | -2(2) | 0 |
| U$_{ij}$ (Cs4) | 96.5(15) | 46.2(10) | 63(9) | 0 | 2(2) | 0 |
| U$_{ij}$ (Br1) | 27.5(7) | 71.1(10) | 55(6) | -3.4(18) | 3.8(14) | 1.0(6) |
| U$_{ij}$ (Br2) | 24.3(7) | 66.7(10) | 74(6) | -3.6(17) | 1.0(13) | 3.2(6) |
| U$_{ij}$ (Br3) | 66.8(15) | 19.7(9) | 159(11) | 0 | -4(2) | 0 |
| U$_{ij}$ (Br4) | 129(2) | 22.3(10) | 41(11) | 0 | -9(3) | 0 |
| U$_{ij}$ (Br5) | 119(2) | 20.5(10) | 93(11) | 0 | -6(3) | 0 |
| U$_{ij}$ (Br6) | 73.3(16) | 21.7(9) | 116(10) | 0 | -4(2) | 0 |
| U$_{ij}$ (Br7) | 65.9(12) | 67.2(13) | 60(9) | -3.9(19) | 9(2) | 4.9(9) |
| U$_{ij}$ (Br8) | 66.4(12) | 66.7(12) | 16(8) | 0.4(17) | 4.6(19) | 4.3(8) |

Space Group: *P*2$_1$/*m*
$a = 11.341(1)$ Å, $b = 11.548(2)$ Å, $c = 11.39(2)$ Å, $\beta = 92.19(2)°$, Dx = 5.164 g/cm³
CsPbBr$_3$, Z=8
Volume = 1490.61 Å³
Measurement Temperature: 298 K
Crystal Dimensions: ~30 mm x ~25 mm x ~25 mm
Wavelength: 0.29521 Å
2θ range for data collection: 3.284° to 28.52°
Index ranges: -16 ≤ h ≤ 17, -16 ≤ k ≤ 18, -9 ≤ l ≤ 6
Total reflections collected: 3747

**Table S3. Continued**

$R_{int}$ =2.63%, $R_{sigma}$=2.46%
Independent reflections: 1953
Twin law: 1 0 0 0 -1 0 0 0 -1 2
Number of fitting parameters: 110
Largest diff. peak/hole: 1.52(Cs4)/-1.20(Br5) e$^-$/Å$^3$
$R_1$ = 3.58 %, $wR_2$ = 10.87 %, Goodness of Fit = 0.997

[*]Atomic displacement parameters $U_{ij}$ (Å$^2$×10$^3$) are in the order $U_{11}$, $U_{22}$, $U_{33}$, $U_{23}$, $U_{13}$, $U_{12}$.

**Table S4.** Structural Parameters at 1.6 GPa ($P2_1/c$ Space Group, Phase 2*)

| Atoms | x(×10⁴) | y(×10⁴) | z(×10⁴) | $U_{eq}$ (Å³×10³) |
|---|---|---|---|---|
| Pb1 | 7504(2) | 43(6) | 6718(3) | 23.8(12) |
| Pb2 | 5000 | 5000 | 5000 | 25.4(18) |
| Pb3 | 7503(2) | 8520(6) | 2915(3) | 18.9(12) |
| Pb4 | 7501(2) | 3552(6) | 5556(3) | 17.4(12) |
| Pb5 | 5140(3) | 216(7) | 6656(3) | 24.6(13) |
| Pb6 | 10000 | 5000 | 5000 | 26.3(18) |
| Pb7 | 10145(3) | 9783(7) | 3337(4) | 32.6(15) |
| Cs1 | 8914(5) | 9880(12) | 1701(6) | 32(2) |
| Cs2 | 6078(5) | 5124(13) | 6708(6) | 36(3) |
| Cs3 | 8590(4) | 1110(12) | 8618(6) | 31(2) |
| Cs4 | 3600(4) | -3901(11) | 6397(6) | 27(2) |
| Cs5 | 6317(4) | 9083(12) | 4809(6) | 31(2) |
| Cs6 | 8675(6) | 9008(15) | 4820(8) | 48(3) |
| Br1 | 7510(6) | 7035(16) | 3957(8) | 24(3) |
| Br2 | 7502(6) | 1850(15) | 4548(8) | 19(3) |
| Br3 | 9778(7) | 6828(18) | 3942(9) | 32(4) |
| Br4 | 10183(6) | 7770(17) | 2371(8) | 26(3) |
| Br5 | 5199(7) | 2270(20) | 7629(10) | 38(4) |
| Br6 | 3724(6) | 70(15) | 6690(8) | 20(3) |
| Br7 | 4793(7) | 3190(20) | 6059(10) | 37(4) |
| Br8 | 8523(7) | -1235(18) | 7423(9) | 32(4) |
| Br9 | 8583(6) | 1400(18) | 6040(9) | 29(3) |
| Br10 | 6466(7) | -1283(18) | 7420(9) | 31(3) |
| Br11 | 6408(6) | 1378(17) | 6035(8) | 26(3) |
| Br12 | 8728(6) | 9932(16) | 3297(8) | 24(3) |
| Br13 | 8632(8) | 5112(19) | 5124(10) | 39(4) |
| Br14 | 7501(7) | 3020(19) | 7355(9) | 34(4) |
| Br15 | 4962(8) | -2000(20) | 5687(11) | 45(4) |
| Br16 | 6375(8) | 5070(20) | 5099(11) | 40(4) |
| Br17 | 10013(10) | 7990(30) | 5658(14) | 62(6) |
| Br18 | 7490(8) | -3070(20) | 6151(11) | 45(5) |

Space Group: $P2_1/c$
$a$ = 20.907(6) Å, $b$ = 8.486(3) Å, $c$ = 23.362(9) Å, $\beta$ = 89.82(3)°, Dx = 5.568 g/cm³
CsPbBr$_3$, Z=24
Volume =4144.79 Å³
Measurement Temperature: 298 K
Crystal Dimensions: ~30 mm x ~25 mm x ~25 mm

**Table S4. Continued**

Wavelength: 0.29521 Å
2θ range for data collection: 2.948° to 28.448°
Index ranges: -28 ≤ h ≤ 31, -12 ≤ k ≤ 10, -26 ≤ l ≤ 24
Total reflections collected: 7292
$R_{int}$ =35.6%, $R_{sigma}$=52.1%
Independent reflections: 5353
Number of fitting parameters: 120
Largest diff. peak/hole: 9.15(Pb6)/-7.80(Pb6) e$^-$/Å$^3$
$R_1$ = 28.6 %, $wR_2$ = 52.8 %, Goodness of Fit = 1.233

---

*Atomic displacement parameters are limited to be isotropic.

**Table S5.** Structural Parameters at Ambient Pressure After Pressure Release ($P2_1/m$, Phase 1)

| Atoms | x(×10⁴) | y(×10⁴) | z(×10⁴) | U$_{eq}$ (Å³×10³) | | |
|---|---|---|---|---|---|---|
| Pb1 | 0 | 5000 | 5000 | 51(3) | | |
| Pb2 | 5000 | 5000 | 0 | 46(3) | | |
| Pb3 | 5000 | 5000 | 5000 | 53(3 | | |
| Pb4 | 0 | 5000 | 0 | 50(3) | | |
| Cs1 | 2534(3) | 7500 | 7411(5) | 100(8 | | |
| Cs2 | 2431(4) | 7500 | 2571(5) | 92(8) | | |
| Cs3 | 7396(5) | 7500 | 2416(5) | 85(7) | | |
| Cs4 | 2496(4) | 2500 | 2438(4) | 88(8) | | |
| Br1 | 2503(3) | 4908(13) | 4862(8) | 171(11) | | |
| Br2 | 2499(3) | 4756(14) | 129(8) | 149(10 | | |
| Br3 | -136(8) | 5141(14) | 2495(3) | 166(13) | | |
| Br4 | 5117(7) | 5091(13) | 2493(3) | 167(13) | | |
| Br5 | 4747(9) | 7500 | 18(7) | 108(4) | | |
| Br6 | 16(17) | 2500 | 4788(16) | 300(30) | | |
| Br7 | -296(11) | 2500 | 153(12) | 179(19) | | |
| Br8 | 5075(15) | 7610(60) | 5106(10) | 210(30) | | |
| U$_{ij}$ (Pb1) | 58.8(19) | 43(10) | 51.2(18) | 1.6(16) | 1.0(12) | 4.5(15) |
| U$_{ij}$ (Pb2) | 57.8(18) | 28(9) | 53.1(17) | 2.7(15) | -2.8(11) | -0.3(14) |
| U$_{ij}$ (Pb3) | 58.2(19) | 50(10) | 50.1(18) | 1.5(16) | 0.0(12) | -1.8(16) |
| U$_{ij}$ (Pb4) | 60.0(19) | 36(10) | 53.5(17) | 3.1(16) | 0.1(12) | 0.3(16) |
| U$_{ij}$ (Cs1) | 80(4) | 130(30) | 91(4) | 0 | 5(3) | 0 |
| U$_{ij}$ (Cs2) | 93(4) | 80(30) | 99(4) | 0 | 5(3) | 0 |
| U$_{ij}$ (Cs3) | 94(4) | 60(20) | 103(4) | 0 | -32(3) | 0 |
| U$_{ij}$ (Cs4) | 133(5) | 60(30) | 74(4) | 0 | -8(3) | 0 |
| U$_{ij}$ (Br1) | 31(3) | 250(30) | 232(11) | -51(9) | 6(4) | -3(4) |
| U$_{ij}$ (Br2) | 33(3) | 160(30) | 254(12) | 25(9) | -2(4) | 8(4) |
| U$_{ij}$ (Br3) | 197(10) | 260(40) | 38(3) | 18(4) | 3(4) | -31(9) |
| U$_{ij}$ (Br4) | 205(10) | 260(40) | 38(3) | 13(4) | 4(4) | -73(10) |
| U$_{ij}$ (Br6) | 290(20) | 430(100) | 188(14) | 0 | 124(16) | 0 |
| U$_{ij}$ (Br7) | 119(7) | 240(60) | 176(10) | 0 | 27(6) | 0 |
| U$_{ij}$ (Br8) | 231(17) | 280(100) | 133(9) | -30(40) | 82(11) | 140(50) |

Space Group: $P2_1/m$
$a = 11.657(3)$ Å, $b = 11.73(2)$ Å, $c = 11.720(3)$ Å, $\beta = 90.02(3)°$, Dx = 4.807 g/cm³
CsPbBr$_3$, Z=8
Volume =1602.55 Å³

**Table S5.** Continued

Measurement Temperature: 298 K
Crystal Dimensions: ~30 mm x ~25 mm x ~25 mm
Wavelength: 0.29521 Å
2θ range for data collection: 3.23° to 28.136°
Index ranges: -18 ≤ h ≤ 17, -6 ≤ k ≤ 9, -19 ≤ l ≤ 17
Total reflections collected:: 3372
$R_{int}$ =7.71%, $R_{sigma}$=10.5%
Twin law: 1 0 0 0 -1 0 0 0 -1 2
Independent reflections:1843
Number of fitting parameters: 110
Largest diff. peak/hole: 3.43(Pb4)/-3.40(Br6) e⁻/Å³
$R_1$ = 16.1 %, $wR_2$ = 44.5 %, Goodness of Fit = 1.325

*Atomic displacement parameters $U_{ij}$ (Å²×10³) are in the order $U_{11}$, $U_{22}$, $U_{33}$, $U_{23}$, $U_{13}$, $U_{12}$. Br5 has isotropic U.

**Table S6.** End-Point Lattice Parameters for Phase 3.

| Pressure (GPa) | a (Å) | b (Å) | c (Å) | β (°) |
|---|---|---|---|---|
| 4.65 | 26.236(8) | 11.034(3) | 26.18(3) | 90.48(7) |
| 10.62 | 25.31(6) | 10.690 (3) | 25.371(8) | 89.91(7) |

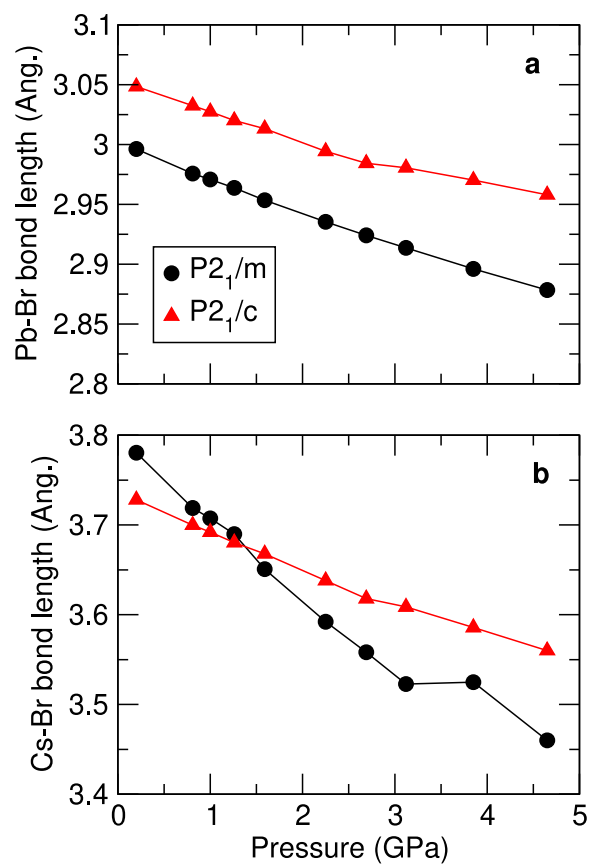

**Figure S14.** (a) Average Pb-Br bond length as a function of pressure for the *P2$_1$/m* and *P2$_1$/c* phases of CsPbBr$_3$. (b) Average Cs-Br bond length as a function of pressure for the *P2$_1$/m* and *P2$_1$/c* phases of CsPbBr$_3$.

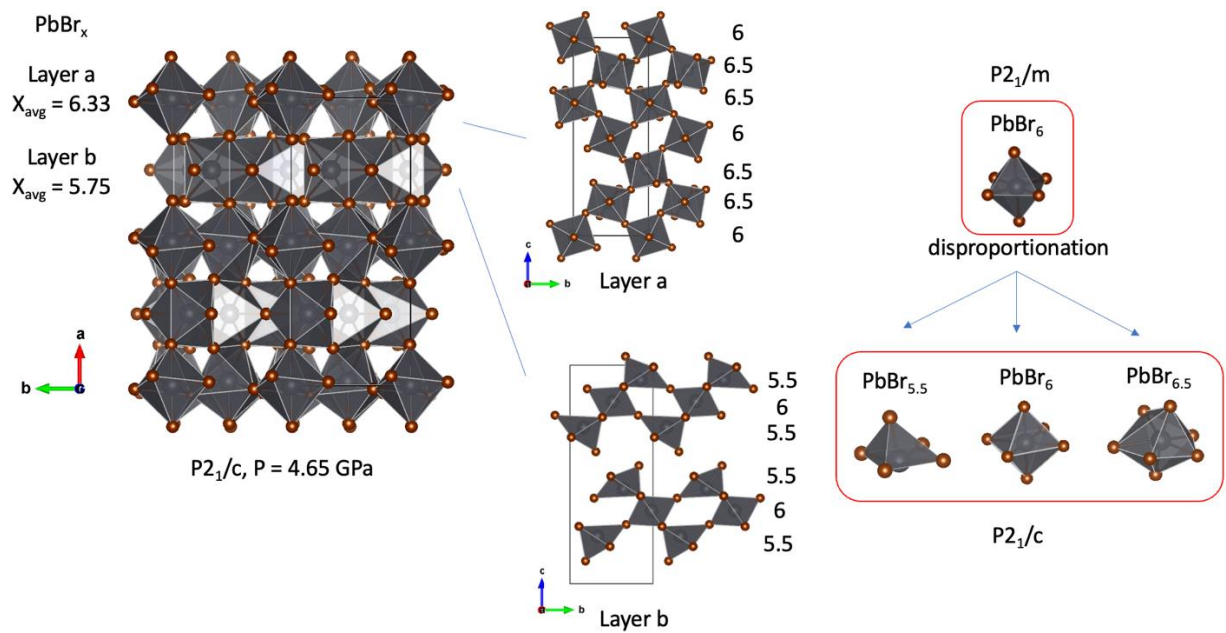

**Figure S15.** Schematic of the PbBr polyhedra arrangement in the *P2₁/c* phase of CsPbBr₃, and the disproportionation of the *P2₁/m* PbBr₆ octahedra across the phase transition.

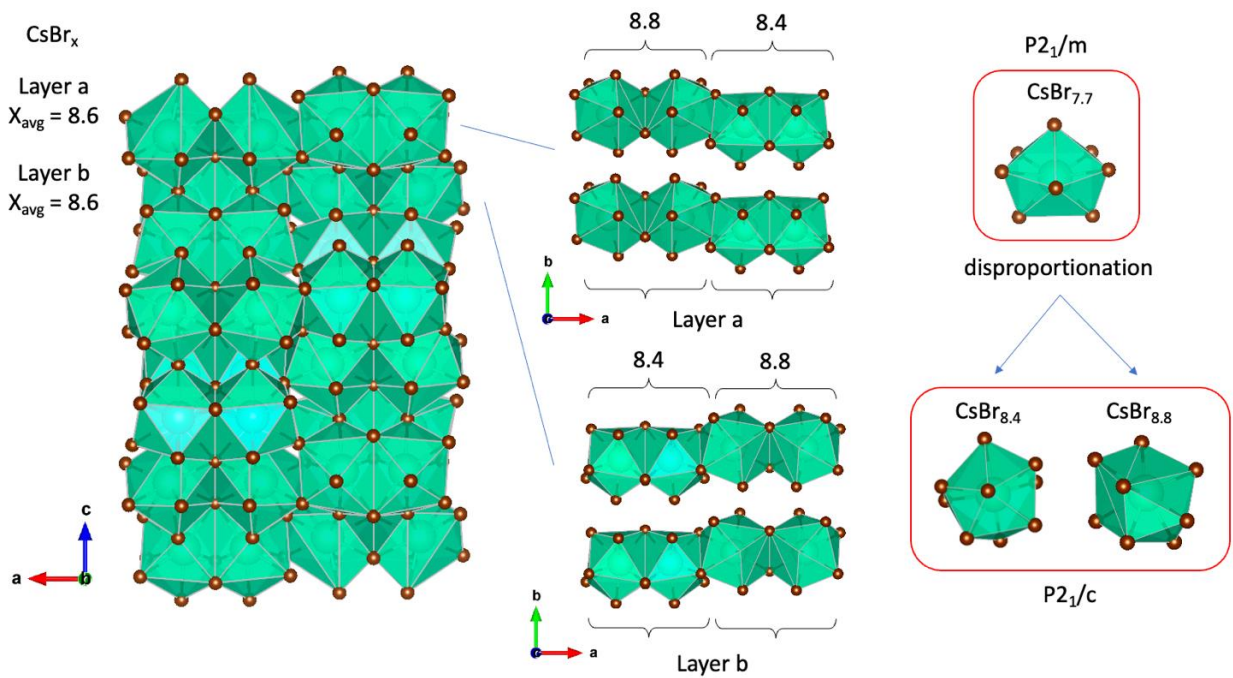

**Figure S16.** Schematic of the CsBr polyhedra arrangement in the *P2₁/c* phase of CsPbBr₃, and the disproportionation of the *P2₁/m* CsBr₈ polyhedra across the phase transition.

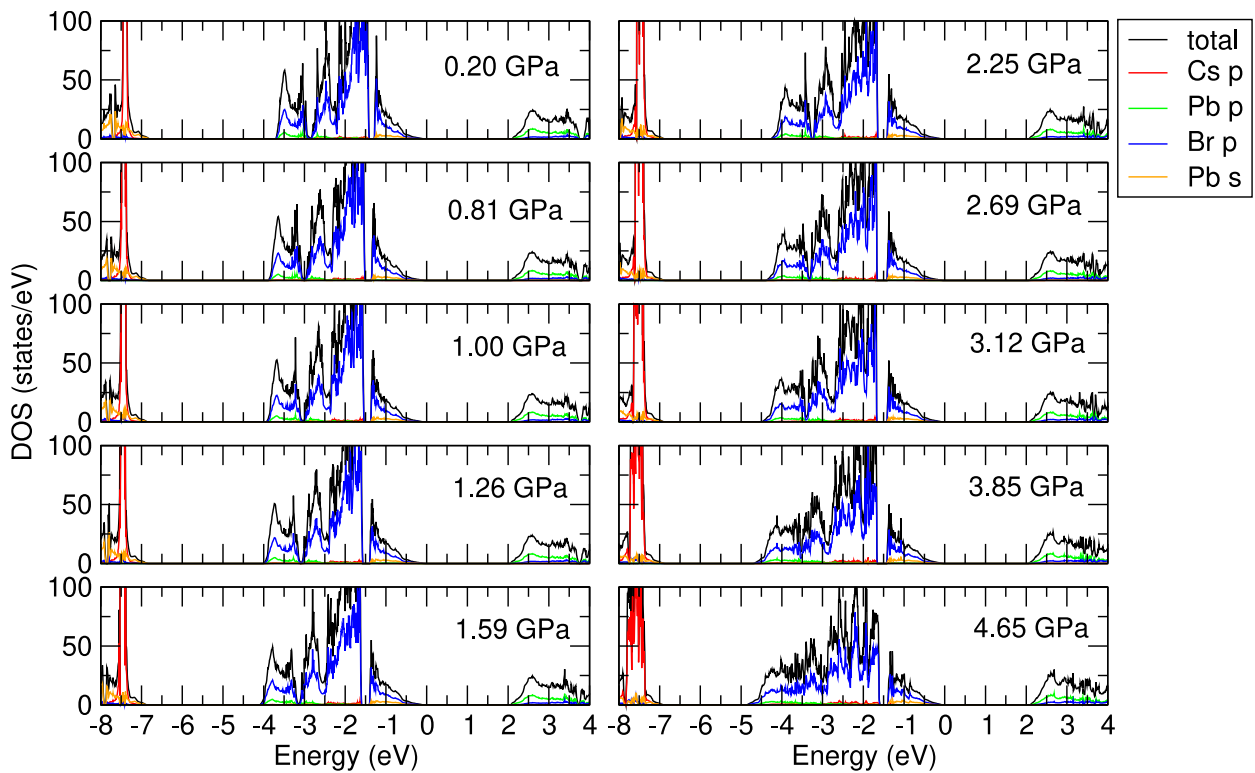

**Figure S17**. Atom resolved density of states for the CsPbBr$_3$ *P2$_1$/m* phase under pressure.

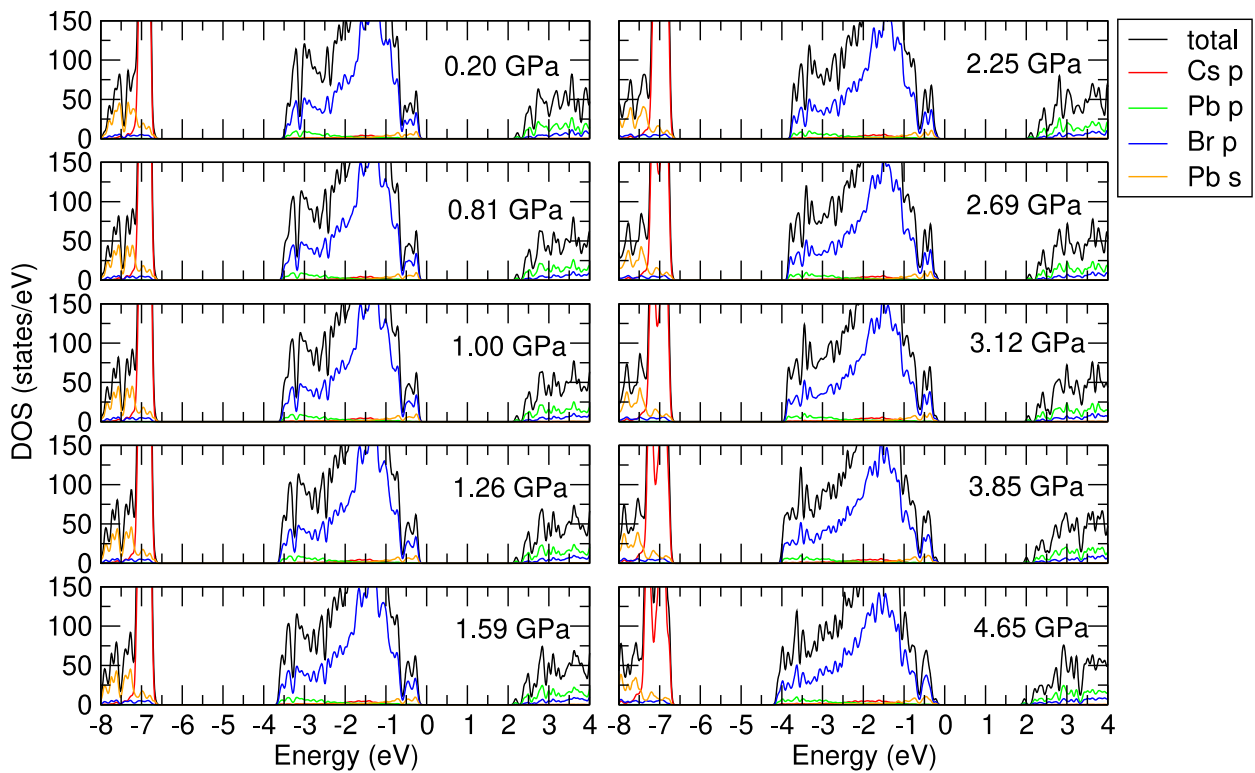

**Figure S18**. Atom resolved density of states for the CsPbBr$_3$ *P2$_1$/c* phase under pressure.

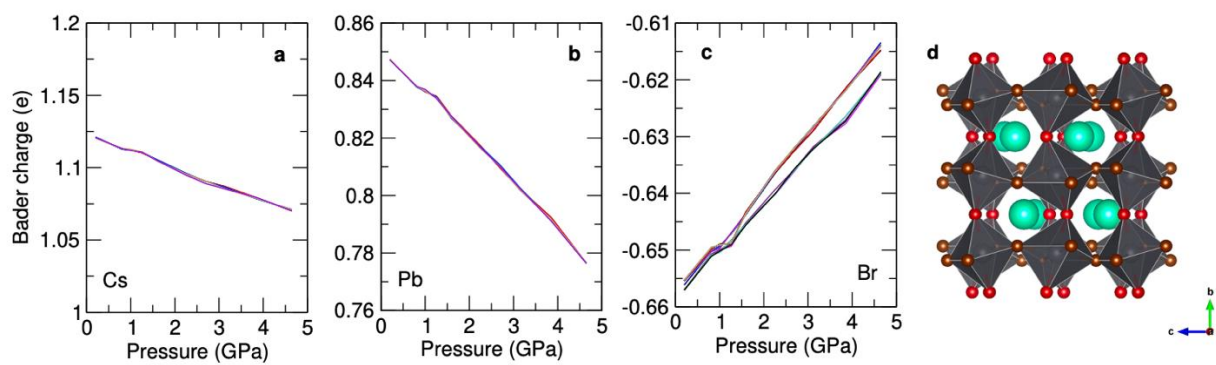

**Figure S19.** Bader charge on (a) Cs atoms, (b) Pb atoms, and (c) Br atoms in the CsPbBr$_3$ $P2_1/m$ phase as a function of pressure. (d) The $P2_1/m$ phase at 4.65 GPa with the Br atoms labeled in red.